# Controlling Hot Electron Spatial and Momentum Distributions in Nanoplasmonic Systems: Volume versus Surface Effects

Jacob Pettine[1,2], Sean M. Meyer[4], Fabio Medeghini[1], Catherine J. Murphy*[4,5], and David J. Nesbitt*[1,2,3]

[1]JILA, University of Colorado Boulder and National Institute of Standards and Technology, Boulder, Colorado 80309, United States

[2]Department of Physics, University of Colorado Boulder, Boulder, Colorado 80309, United States

[3]Department of Chemistry, University of Colorado Boulder, Boulder, Colorado 80309, United States

[4]Department of Chemistry, University of Illinois at Urbana-Champaign, Urbana, Illinois 61801, United States

[5]Materials Research Laboratory, University of Illinois at Urbana-Champaign, Urbana, Illinois 61801, United States

*Correspondence and requests for materials should be addressed to C.J.M. (murphycj@illinois.edu) and D.J.N. (djn@jila.colorado.edu).

# Abstract

Hot carrier spatial and momentum distributions in nanoplasmonic systems depend sensitively on the optical excitation parameters and nanoscale geometry, which therefore determine the efficiency and functionality of plasmon-enhanced catalysts, photovoltaics, and nanocathodes. A growing appreciation over the past decade for the distinction between volume- and surface-mediated photoexcitation and electron emission from such systems has underscored the need for direct mechanistic insight and quantification of these two processes. Toward this end, we use angle-resolved photoelectron velocity mapping to directly distinguish volume and surface contributions to nanoplasmonic hot electron emission from gold nanorods as a function of aspect ratio, down to the spherical limit. Nanorods excited along their longitudinal surface plasmon axis exhibit surprising transverse photoemission distributions due to the dominant volume excitation mechanisms, as reproduced via ballistic Monte Carlo modelling. We further demonstrate a screening-induced transition from volume (transverse) to surface (longitudinal) photoemission with red detuning of the excitation laser and determine the relative cross-sections of the two mechanisms via combined volume and surface multiphoton photoemission modelling. Based on these results, we are able to identify geometry- and material-specific contributions to the photoemission cross-sections and offer general principles for designing nanoplasmonic systems to control hot electron excitation and emission distributions.

# Introduction

Observations of electron emission from solids under ultraviolet or x-ray irradiation have been essential to early developments in quantum mechanics and modern insights into electronic structure and physical properties of materials (1). Of particular recent interest has been the elucidation of electron emission dynamics from nanoscale materials, with a special emphasis on plasmonic metal nanoparticles and nanostructures (2-9). The relatively low-energy visible resonances and extraordinary optical field concentration in these systems has served to revitalize the century-old problem (10-13) of distinguishing between (i) electron excitation throughout the volume of the material, followed by ballistic transport and escape, versus (ii) excitation and emission directly at the surface (12). Understanding these mechanisms will unlock opportunities for nanoscale control over hot carriers in emerging plasmonic photocatalytic (13-15), photovoltaic (12, 16), and nanocathode applications (7, 8, 17), among others.

At the heart of this issue is the negligible linear momenta of the incoming photons compared with the outgoing electrons. Momentum conservation thus demands electron scattering with a massive third body during photoexcitation and emission. Photoemission via volume excitation is dominated by scattering with the periodic lattice potential when the transition is energetically allowed, but visible plasmonic excitation is often below the relevant interband threshold and instead involves interactions with phonons, defects, impurities, or other electrons. By contrast, surface photoexcitation and emission pathways arise due to the translational symmetry breaking at an interface and thus involve scattering with the surface potential barrier, including contributions from the electromagnetic field variation, localized surface states, and the evanescent external decay of internally-delocalized Bloch wavefunctions (11). For nanoscale systems with ≲ 20 nm dimensions, intraband excitation mediated by geometrical confinement can also become significant (18-20).

Given the strong and highly-spatially-varying electric field enhancements in nanoplasmonic systems, volume vs. surface photoexcitation will often lead to disparate spatial and angular photocurrent distributions, which can be harnessed in plasmonic hot carrier devices. Hot carrier catalysts, for instance, already exhibit high reaction efficiencies and product specificity compared with thermally-activated processes (14, 15), as demonstrated via $CO_2$ conversion (21) and $H_2O$ splitting (13, 22) for solar fuel production. However, further enhancements in catalytic activity and device functionality can be achieved by controlling the hot carrier spatial and momentum degrees of freedom to compliment anisotropic coatings (16, 22-24) or even to introduce nanometer site selectivity (6, 25). Similar opportunities exist in broadband photodetection (26, 27) and solar energy conversion (28, 29). Plasmonic nanoparticles and nanostructure arrays also serve as bright photocathodes (3, 30), with possibilities for optically-controlled current directionality (8) in terahertz nanoelectronics (17) and femtosecond electron imaging (31) applications. While internal hot electron emission at metal-molecule (13) or metal-semiconductor (12, 22, 23) interfaces is often classified as either "direct" excitation at the surface or "indirect" transfer following volume excitation (14, 15, 29, 32), these map directly onto the distinction between surface and volume mechanisms in external (metal-vacuum) photoemission. A deeper understanding and control of volume vs. surface effects is therefore essential to optimizing hot carrier device performance, regardless of the application or collection medium.

The primary focus of the present work is to distinguish volume vs. surface photoemission pathways in plasmonic nanoparticles by their dramatically different photoelectron momentum distributions. We begin by showing that resonant longitudinal excitation of gold nanorods leads to transverse (orthogonal) multiphoton photoemission (MPPE) distributions due to the dominance of volume excitation mechanisms. We then demonstrate a novel transition from volume (transverse) to surface (longitudinal) MPPE that occurs with red detuning of the excitation laser, resulting from the enhanced metallic screening of internal electric fields at lower frequencies. Detailed modelling of the volume and surface photoemission distributions reproduces all of these effects and allows us to quantitatively distinguish the MPPE cross-sections. Most importantly, we demonstrate that the relative surface vs. volume MPPE depends primarily on plasmonic field enhancements, which can be modelled via classical finite element, finite difference, or other methods. Although measurements are performed in the 2-, 3-, and 4-photon regimes to overcome the ~4.25 eV gold nanoparticle work function with visible excitation frequencies, the analysis and conclusions are completely general with respect to process order and should therefore remain valid down to the linear regime for 1-photon applications. Finally, these results allow us to offer general design principles for engineering volume and surface processes and thereby controlling hot electron excitation and emission distributions in designer nanoplasmonic systems.

## Results

**Transverse Volume Photoemission from Longitudinally-Excited Nanorods**

Strong electric near-field enhancements are generated at the tips of gold nanorods where conduction electrons collectively pile up during longitudinal surface plasmon resonance (SPR) oscillations. At the same time, appreciable field enhancements are also generated within the metal volume. High densities of hot carriers may therefore be excited at nanorod tips via surface excitation mechanisms or inside the nanorod via volume excitation mechanisms. The corresponding photoemission pathways are illustrated in Fig. 1A, along with the scanning photoelectron imaging microscopy (SPIM) experimental configuration (*Methods*). Different photoemission angular distributions are expected for the two mechanisms (33), depending upon the nanoparticle geometry and the electric near-field distribution of the excited plasmon mode. This provides a direct means of identifying volume and surface photoemission contributions via single-particle, angle-resolved photoelectron velocity mapping.

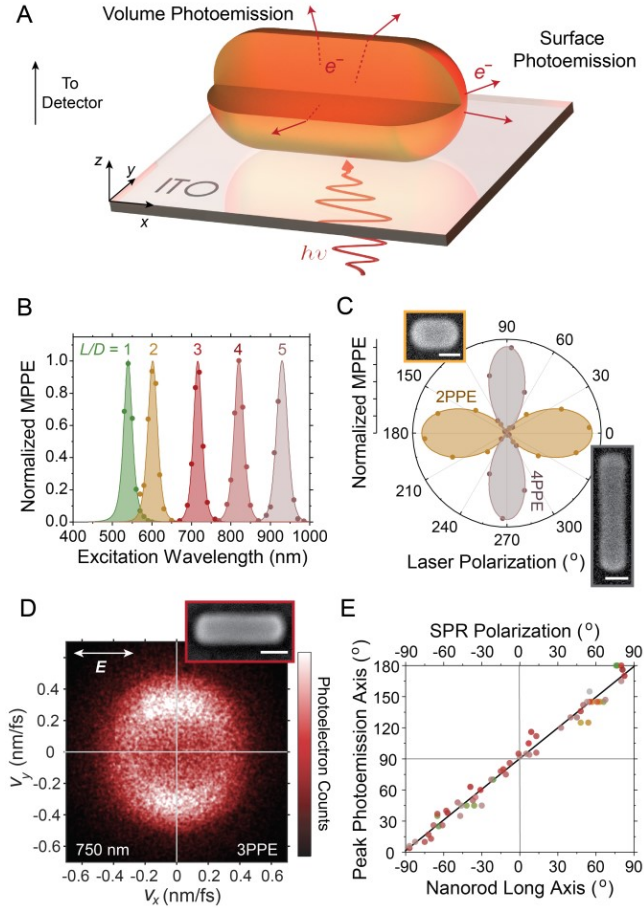

**Fig. 1.** Characterization of nanorod surface plasmon resonance photoemission properties. (*A*) Configuration of scanning photoelectron imaging microscopy experiments, illustrating volume and surface emission from an illuminated gold nanorod (with a quarter section removed to show the volume excitation). (*B*) Multiphoton photoemission spectra measured for nanorods of various aspect ratios, fit to nonlinear Lorentzian profiles. (*C*) Signal dependence on linear laser polarization in the azimuthal ($xy$) plane, fit to $\cos^{2n}(\theta - \theta_{\text{rod}})$, shown with scanning electron micrographs of the correlated nanorods ($L/D = 1.5$, 2PPE and $L/D = 4.5$, 4PPE). Laser polarization $\theta = 0°$ is along the $x$ axis and $\theta = 90°$ is along the $y$ axis. (*D*) Photoelectron velocity map collected at the longitudinal resonance of the correlated $L/D = 3.2$ nanorod in the inset, exhibiting transverse photoemission. (*E*) Summary of photoemission directionality for all nanorods of various aspect ratios and spatial orientations, with transverse peak photoemission observed in every case. Data colors (here and elsewhere) are mapped to the corresponding wavelength. Scale bars: 20 nm.

Photoemission spectra are shown in Fig. 1B for nanorods of similar diameter, $D = 21 \pm 4$ nm (*SI Appendix*, Fig. S1), but different lengths and thus aspect ratios ($L/D$), illustrating the anticipated linear increase of SPR wavelength with nanorod aspect ratio for $L/D \gtrsim 2$ (34, 35) as summarized in the *SI Appendix*, Fig. S2. For the gold nanorod work function around 4.25 eV, electrons must absorb multiple photons to overcome the surface potential barrier, with nanorod resonances ranging from 950 nm (1.3 eV photon energy, $n = 4$ photons) at $L/D = 5$ down to the spherical limit of 540 nm (2.3 eV photon energy, $n = 2$ photons) at $L/D = 1$. Measurements of the photoemission dependence on linear laser polarization ($\theta$) in the azimuthal plane (Fig. 1C) show clearly-defined longitudinal resonances along the long nanorod axes, noticeably narrowing in the 4-photon regime ($L/D \sim 4.5$) relative to the 2-photon regime ($L/D \sim 1.5$) due to the $n\text{PPE} \propto E^{2n}\cos^{2n}(\theta - \theta_{\text{rod}})$ dependence of the field projection along the nanorod axis.

While longitudinal electron emission outward from the highly field-enhanced nanorod tips has been observed in the strong-field tunneling (7, 36) and transitional regimes (5), we surprisingly find that weak-field MPPE is predominantly transverse (orthogonal) to the resonantly-excited longitudinal nanorod axis, as demonstrated in Fig. 1D. In other words, electrons are evidently emitted from the sides rather than the tips of the nanorods. We observe this transverse photoemission for every nanorod investigated (Fig. 1E), irrespective of aspect ratio ($L/D = 1.5\text{-}5$), surface ligands (*SI Appendix*, Fig. S5), size ($D = 10\text{-}40$ nm, *SI Appendix*, Fig. S5), or corresponding size-dependent differences in faceting (37). Such observations provide a strong initial indication that the transverse photoemission arises due to volume-mediated hot electron excitation for the nanorod geometry. In such a mechanism, one expects hot electrons excited throughout the nanorod to escape ballistically from the cylindrical body with a predominantly transverse distribution and from the hemispherical tips with a nearly isotropic distribution. As a starting point, if we assume uniform excitation throughout the nanorod volume and a short inelastic mean free path ($\lambda_{\text{inel}} \ll D$), the relative side versus tip contributions can be approximately estimated by the ratio of corresponding surface areas, which works out to be $S_\text{side}/S_\text{tip} = L/D - 1$. We therefore expect the photoemission distribution to become isotropic as $L/D \to 1$ ($S_\text{side} \ll S_\text{tip}$) or to become increasingly dominated by transverse contributions as $L/D \to \infty$ ($S_\text{side} \gg S_\text{tip}$).

To show definitively that the photoemitted electrons primarily originate within the volume of the resonantly-excited nanorods, we explore the photoemission distributions as a function of nanorod aspect ratio. As expected, the 2D velocity maps in Fig. 2A demonstrate that the photoemission angular distributions become more isotropic with decreasing aspect ratio and completely isotropic in the spherical limit. The radially-integrated angular distributions in Fig. 2C can be characterized by an angular contrast, $AC$, defined as

$$AC = \frac{\langle \| \text{ counts} \rangle - \langle \perp \text{ counts} \rangle}{\langle \| \text{ counts} \rangle + \langle \perp \text{ counts} \rangle}, \qquad [1]$$

where the brackets denote averaging (within $\pm 2°$) over the two longitudinal (0° and 180°) and transverse (90° and 270°) directions on the 2D velocity maps. This definition of the angular contrast provides a model-independent metric of how transverse ($AC \to -1$) or longitudinal ($AC \to 1$) a given distribution is. Angular contrast values are summarized as a function of aspect ratio in Fig. 2E, where $AC$ clearly becomes more negative (transverse) as the nanorod aspect ratio increases, and goes to 0 in the isotropic spherical limit, as expected for the volume photoemission mechanism. We note that the angular contrast, $AC$, and the photoemission mechanisms described here are not to be confused with molecular photoionization and the corresponding anisotropy parameter, $\beta$ (38).

A summary of nonlinear process order, $n$ (where $n\text{PPE} \propto I_0^n$), from intensity-dependence measurements performed on resonance for each nanorod studied (*SI Appendix*, Fig. S3) is shown in Fig. 2D. The results show a clear sequential transition from 2PPE to 3PPE to 4PPE with increasing SPR wavelength and linearly-correlated increasing aspect ratio (Fig. 1B and *SI Appendix*, Fig. S2). Non-integer process orders in the transition regions arise naturally in the power-law fits and reflect the weighted sum of the two contributing process orders (*SI Appendix*, Section 2) (8). The well-defined integer process orders and transitions verify that the present studies are performed in the perturbative MPPE regime rather than the optical field emission or thermionic regimes (see *SI Appendix*, Section 2 for an extended discussion). We also note that no sudden changes in photoemission angular distributions or corresponding angular contrast values are observed in the transition regions between process orders. The physics is qualitatively the same at all nonlinear orders studied herein ($n = $ 2-4), simply with a different total excitation energy $n\hbar\omega$ and nonlinear absorption cross-sections. We thus anticipate that the present techniques can be extrapolated from the multiphoton regimes down to the linear emission regime for systems with lower hot electron emission barriers, including metal-semiconductor and metal-molecule junctions.

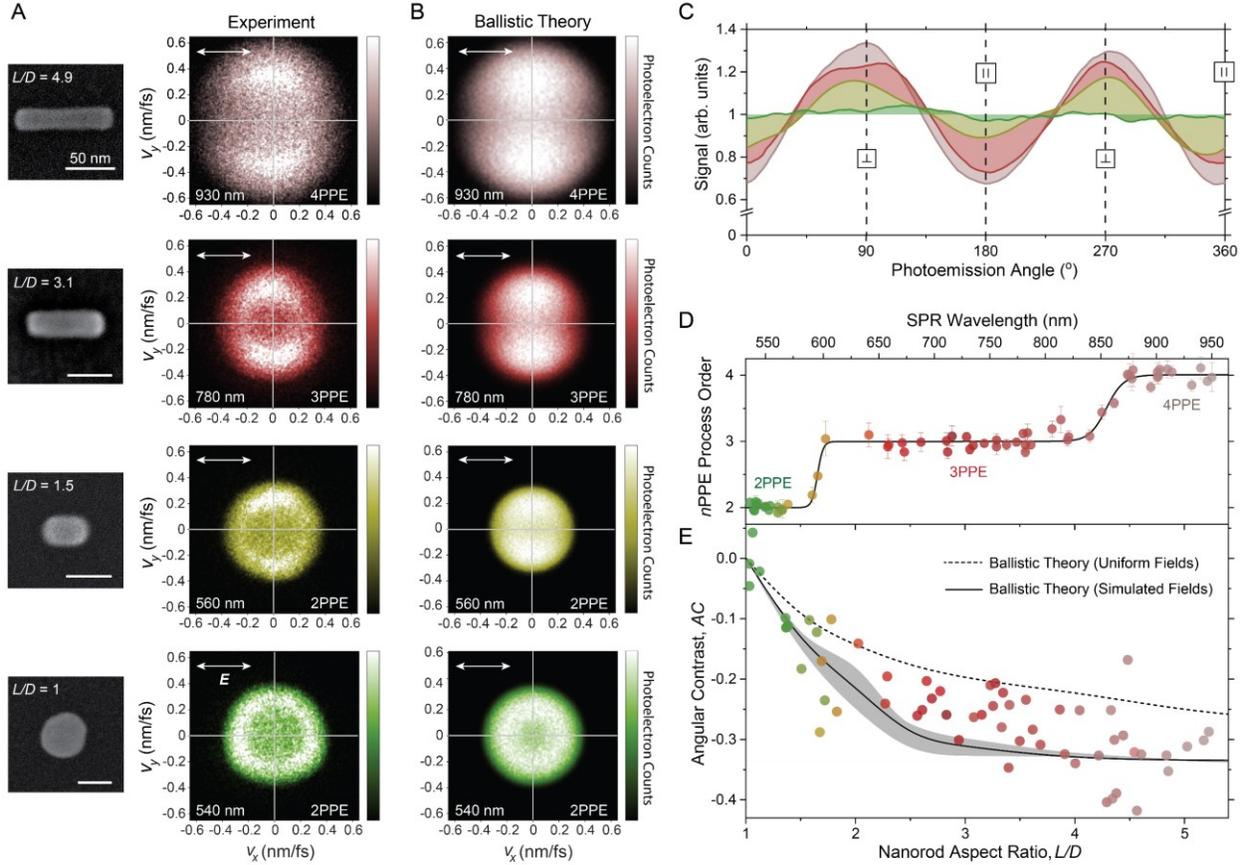

**Fig. 2.** Volume photoemission distributions as a function of aspect ratio. (A) Correlated electron micrographs and velocity maps for a series of nanorods excited at their longitudinal surface plasmon resonances (SPRs), down to the spherical limit. Nanospheres of larger $D = 70$ nm were studied compared with nanorods ($D = 20$ nm) for sufficient signal levels, but this is neither expected nor observed to affect their photoemission properties. Scale bars: 50 nm. (B) Velocity distributions for the same nanorods as in (A), modelled using the ballistic (3-step) Monte Carlo theory. (C) Radially-integrated angular distributions from the velocity maps in (A) with transverse ($\perp$) and longitudinal ($\parallel$) directions indicated and 0° along the $x$ (i.e. $v_x$) polarization axis. (D) Process order summary of $n$-photon photoemission measured via power-law intensity-dependence fits (SI Appendix, Fig. S4 – error bars are standard errors of the fits) for all investigated nanorods for longitudinal SPR excitation, shown with overall sigmoidal fits. The SPR wavelength axis is linearly-mapped (except around $L/D = 1$) to the aspect ratio (SI Appendix, Fig. S2). (E) Photoemission angular contrast of all investigated nanorods for longitudinal SPR excitation, becoming more negative (transverse) with increasing aspect ratio and isotropic in the spherical limit. Ballistic Monte Carlo theory curves shown for uniform and nonuniform (finite-element-simulated) internal fields. The error bounds shown for the simulated fields case assume an inelastic mean free path between 7 nm (top) and 5 nm (bottom), with the primary curve calculated at 6 nm.

To understand the volume MPPE distributions in further detail, we implement a Monte Carlo algorithm within a ballistic, three-step photoemission model (39). In this model, electrons are first excited to a randomly-selected (isotropic, Fermi-Dirac-weighted) vector momentum at a random point within the nanorod volume, weighted by the nonlinear internal field enhancement ($|E/E_0|^{2n}$) determined via electrodynamic finite element simulations (*Methods*). The hot electrons then travel ballistically over some distance ($d$) to the surface with an exponential survival probability for inelastic hot-cold electron-electron scattering, $e^{-d/\lambda_{\text{inel}}}$. The inelastic mean free path, $\lambda_{\text{inel}} \approx 6$ nm, is approximately constant over the narrow threshold energy range of interest (19, 40) and we account for the possibility of electrons surviving a single inelastic scattering event with sufficient momentum to subsequently escape, although this contribution is found to be negligible (*SI Appendix*, Section 4). Finally, hot electrons that reach the surface are transmitted into the surrounding medium with unity probability if they have sufficient surface-normal momentum to overcome the surface potential barrier and are otherwise reflected and lost. Trajectories following surface reflection could be readily incorporated into the modelling and may be relevant for particle dimensions similar to or less than the inelastic mean free path (41, 42), but are safely neglected here. While other integration methods have been used to solve for the spatial distributions of emitted hot electrons (41, 43), the Monte Carlo method provides a computationally efficient means of integrating over all hot electron trajectories for arbitrary nanoparticle geometries (6, 42, 44). Further details of the ballistic Monte Carlo method are described in *SI Appendix*, Section 4.

The calculated Monte Carlo velocity maps in Fig. 2B and angular contrasts in Fig. 2E recapitulate the experimental angular distributions reasonably well. Small longitudinal surface photoemission contributions are likely the cause of the slightly less negative experimental angular contrast values compared with the calculations, as examined in the next section. Except for near the Fermi edge, which is in good agreement due to the experimental determination of the $4.25 \pm 0.1$ eV work function from the process order transitions (Fig. 2D) and from the velocity maps, the radial dependence of the photoemission distribution is evidently not reproduced as well by the approximation of constant joint density of states and energy-independent excitation matrix elements utilized in the Monte Carlo modelling. Further details of the nascent hot electron distribution could be incorporated into the modelling, but this would require a significantly more detailed *ab initio* treatment of the material band structure and a variety of multiphoton volume excitation channels, including direct versus indirect (phonon-mediated) transitions in coherent and incoherent channels. Such *ab initio* calculations have been demonstrated by Narang and colleagues for two-photon excitation (45) but are beyond the scope of the present work.

**Controlling Volume vs. Surface Photoemission**

For further insight into the roles of both volume and surface photoemission processes, we simulate the electric near-field distributions as a function of excitation frequency (and thus detuning from SPR) in Fig. 3A. Unlike the well-known uniform internal fields for ellipsoids (46), internal fields for nanorods are stronger near the center due to a hemispherical surface charge distribution at the tips (dipolar in the spherical limit) with field vectors that destructively cancel within the tip regions but add constructively near the nanorod center. Such central concentration of the field distributions leads to more centralized excitation of hot electrons, which subsequently escape predominantly from the sides of the nanorods rather than the tips due to the limiting inelastic mean free path. This results in more negative (transverse) angular contrast values compared with uniform excitation throughout the nanorod body, as seen by comparing the theoretical curves Fig. 2E. It should be noted that this is a very different phenomenon from centralized heating effects that have been observed due to higher electron kinetic energy in the center of the nanorod during plasmon oscillations (7), as thermal effects are negligible in these relatively low-intensity (peak $I_0 \approx 0.05$ GW cm$^{-2}$) perturbative studies (*SI Appendix*, Fig. S4).

While the overall field enhancement is strongest on the plasmon resonance, the relative surface field contribution increases with red detuning due to enhanced screening of the internal fields at lower frequencies by the conduction electrons. The ratio of these nonlinear surface and volume field enhancement integrals (proportional to the nonlinear photoemission cross-section ratio, $\sigma_S^{(n\text{PPE})}/\sigma_V^{(n\text{PPE})}$) is summarized for different aspect ratios in Fig. 3B. The total multiphoton photoemission rate can be written as

$$\text{MPPE} = \sum_n \left( \sigma_S^{(n\text{PPE})}(\omega) + \sigma_V^{(n\text{PPE})}(\omega) \right) I_0^n, \qquad [2]$$

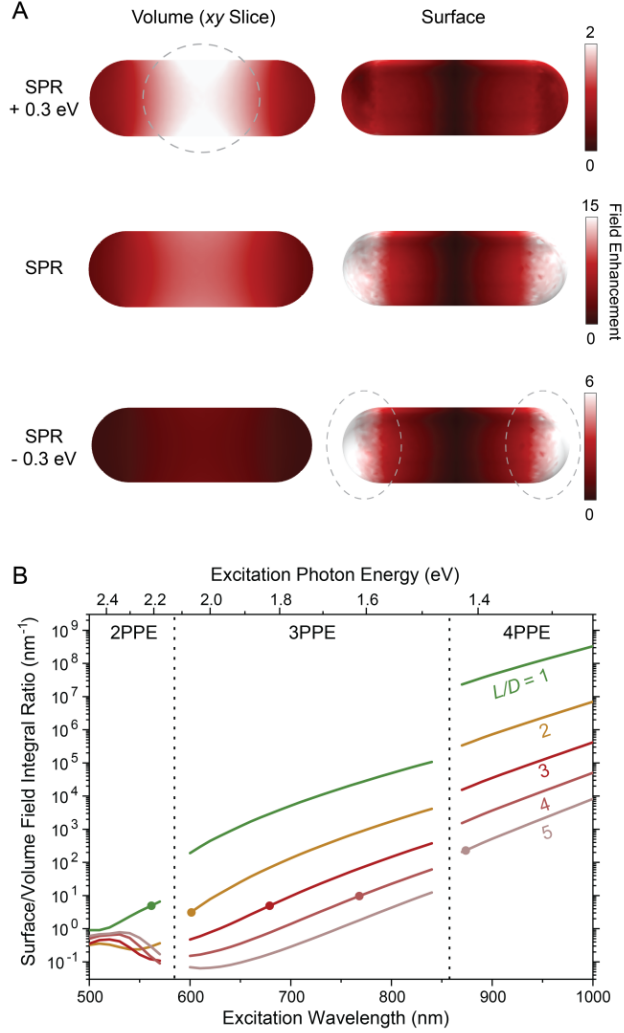

**Fig. 3.** Finite element simulations of surface and volume plasmonic field enhancements. (*A*) Volume ($|E/E_0|$) and surface ($|E_\perp/E_0|$) field enhancement maps for $L/D = 3$ nanorod viewed in the $xy$ plane (ligand layer and ITO substrate accounted for but not shown; see *Methods*). Volume fields are dominant at higher excitation energies (e.g. SPR + 0.3 eV) while surface fields are dominant at lower excitation energies (e.g. SPR − 0.3 eV). (*B*) Ratio of the nonlinear surface field integral (as in Eq. 3A) to the nonlinear volume field integral (as in Eq. 3B) for a series of nanorod aspect ratios across the $n = 2$-$4$ spectral regions. Points indicate the calculated surface plasmon resonance wavelengths. The fluctuations in the 2PPE regime are primarily due to the onset of $5d$-band absorption and dispersion.

where the surface and volume $n$PPE cross-sections are given by

$$\sigma_S^{(n\text{PPE})}(\omega) = c_S^{(n)}(\omega)\,\eta_S^{(n)}(\omega) \int |E_\perp(\mathbf{r},\omega)/E_0|^{2n} dS \quad [3A]$$

and

$$\sigma_V^{(n\text{PPE})}(\omega) = c_V^{(n)}(\omega)\,\eta_V^{(n)}(\omega) \int |E(\mathbf{r},\omega)/E_0|^{2n} dV, \quad [3B]$$

respectively. The $c_S^{(n)}$ and $c_V^{(n)}$ factors are the surface and volume nonlinear absorption densities and $\eta_S^{(n)}$ and $\eta_V^{(n)}$ are the emission quantum yields for a given process order. Strictly speaking, the volume emission quantum yield in Eq. 3B is actually a volume-averaged quantity,

$$\eta_V^{(n)}(\omega) = \frac{\int \eta_V^{(n)}(\mathbf{r}, \omega)|E(\mathbf{r}, \omega)|^{2n} dV}{\int |E(\mathbf{r}, \omega)|^{2n} dV},$$

which therefore depends on the geometry and can be determined via ballistic Monte Carlo modelling using the simulated fields. Typical values for $\eta_V^{(n)}$ for nanorods are between 0.1% and 1%, as quantified via full nanorod surface maps in *SI Appendix*, Fig. S8, although this depends strongly on the frequency (47) and may also be substantially enhanced by surface roughness (48). By comparison, $\eta_S^{(n)}$ only depends on generic properties of a locally-flat surface potential barrier and is therefore not geometry-specific. As a result, the only geometry-specific quantities are the field enhancement integrals in Eqs. 3A and 3B (which encode the plasmonic response of the system) and $\eta_V^{(n)}$, which can be determined via finite element and Monte Carlo modelling, respectively. The remaining $c_S^{(n)} \eta_S^{(n)}$ and $c_V^{(n)}$ values are properties only of the material and/or interface. Thus, if these material quantities can be characterized experimentally, then $\sigma_S^{(nPPE)}$ and $\sigma_V^{(nPPE)}$ are fully determined, permitting the surface and volume contributions to be predicted for arbitrary nanoscale or even macroscopic geometries.

Toward the goal of determining $\sigma_S^{(nPPE)}$ and $\sigma_V^{(nPPE)}$ separately, we utilize photoelectron velocity mapping to distinguish the volume and surface distributions. Considering the dramatic change in the relative nonlinear surface vs. volume field enhancements with detuning (see Fig. 3B), the photoemission is expected to transition from the transverse volume-dominated distribution ($AC < 0$) always observed around the SPR to a longitudinal surface-dominated distribution ($AC > 0$) with only modest red detuning. This is precisely what is observed in Fig. 4A, with the angular contrast summarized in Fig. 4C becoming positive around $\Delta \hbar \omega \approx -0.25$ eV detuning. Unlike transverse volume emission, the longitudinal surface emission is often asymmetric due to amplification of any tip field asymmetry (i.e. due to slight tip curvature differences) by the nonlinear process. For example, a 10% difference in the tip fields results in a factor of two difference in 4PPE rates ($\propto |E/E_0|^8$). To take such asymmetry into account, the angular contrast values are obtained by averaging over both longitudinal directions (Eq. 1). The measured angular contrast values for 9 sample nanorods resonant between 700-800 nm ($L/D = 2.5$-$3.5$) are summarized in Fig. 4C, which all display very similar behaviors with detuning. Notably, the range of detuning values crossing $AC = 0$ is about a factor of three narrower than the spread in absolute energies arising from the spread in the SPR energies due to sample heterogeneity, as explicitly shown in *SI Appendix*, Fig. S9. This indicates that detuning is the predominant cause of the transition from volume- to surface-dominated photoemission for a specific geometry, with no sudden change observed for the 3PPE-4PPE process order transition at a specific excitation energy. Furthermore, intensity-dependence measurements verify that the surface emission at −0.4 eV red detuning remains in the multiphoton regime rather than the strong-field regime (S*I Appendix*, Fig. S10). Higher input intensities are utilized at −0.4 eV red detuning ($I_0 \approx 2$ GW cm$^{-2}$ versus 0.05 GW cm$^{-2}$ on resonance), but this simply maintains similar signal levels by compensating for the drop in the plasmonic field enhancement off resonance.

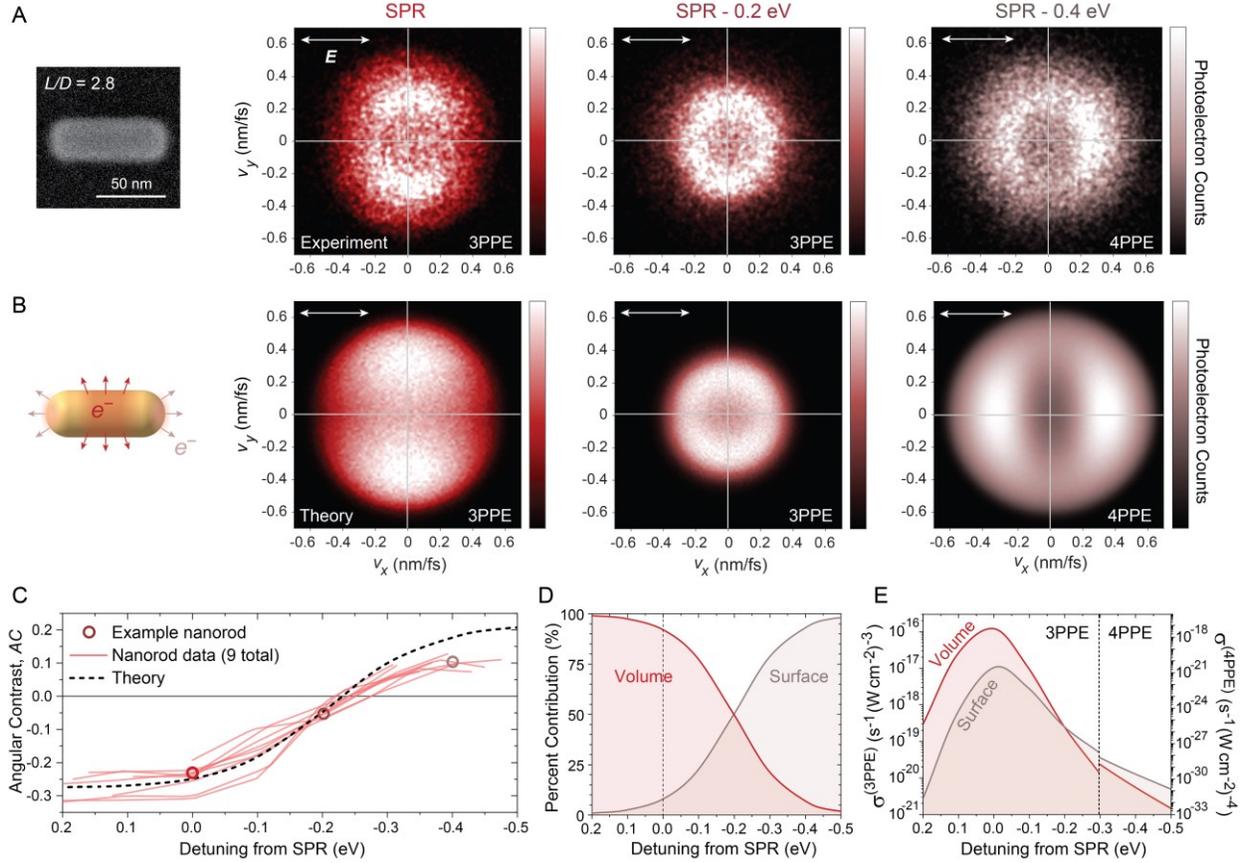

**Fig. 4.** Transition from volume to surface photoemission with red detuning. (A) Experimental and (B) theoretical velocity maps at various detunings from the SPR of the correlated example nanorod, showing the transition from volume (transverse) to surface (longitudinal) photoemission. (C) Summary of angular photoemission contrast values for 9 nanorods, with data points for the example rod in (A) and the theoretical curve from (B). All nanorods display consistent behavior and transition into the surface regime ($AC > 0$) around $-0.25$ eV detuning. (D) Relative volume and surface contributions determined from the fit of $a^{(n)}$ in (C). (E) Volume and surface $n$PPE cross-sections for the example nanorod in (A).

To help theoretically characterize the relative surface and volume contributions to the observed photoelectron distributions, we model the 3D photoemission distributions (and 2D projections) for each mechanism, taking the correlated nanorod geometry into account (Fig. 4B). Volume photoemission distributions are again modelled via the ballistic Monte Carlo method, while the surface MPPE theory developed by Yalunin and coworkers (49) is implemented to model the surface distributions. The implementation of this surface MPPE theory to arbitrary nanoparticle systems is described in the *SI Appendix*, Section 5 and in detail in a previous work (8).

The relative weighting between calculated surface and volume contributions, $\sigma_S^{(n\text{PPE})}/\sigma_V^{(n\text{PPE})}$, depends on the ratio of the field integrals (Fig. 3B) as well as the primarily material-specific (geometry-independent) coefficients in Eqs. 3A and 3B. Since we are working in the threshold photoemission regime with excess photoelectron kinetic energies < 2 eV and approximating a constant density of states and excitation matrix elements in this narrow energy range, the frequency dependence of both surface and volume coefficients can be expected to obey Fowler's law (47), $c_S^{(n)}(\omega)\eta_S^{(n)}(\omega) \propto c_V^{(n)}(\omega)\eta_V^{(n)}(\omega) \propto (n\hbar\omega - \phi)^2$, thus cancelling out in the $\sigma_S^{(n\text{PPE})}/\sigma_V^{(n\text{PPE})}$ ratio. As a result, all of the frequency dependence in $\sigma_S^{(n\text{PPE})}/\sigma_V^{(n\text{PPE})}$ is contained within the ratio of the nonlinear field integrals, scaled by a single frequency-independent prefactor

$$a^{(n)} = \frac{c_S^{(n)}(\omega)\eta_S^{(n)}(\omega)}{c_V^{(n)}(\omega)\eta_V^{(n)}(\omega)}.$$

Thus, the role of experiment in determining the surface and volume photoemission contributions has been reduced to determination of this single $a^{(n)}$ parameter for a given process order, $n$.

Weighting the modelled surface and volume photoemission distributions by the field integral ratio from Fig. 3B, we find that $a^{(3)} \approx 7.5$ pm ($\pm 50\%$) yields the best agreement with the experimental angular contrast as a function of detuning (Fig. 4C). The relative surface and volume contributions are now quantified, as shown in Fig. 4D, which indicates that volume processes account for 90% of the total photoemission on resonance. As suggested earlier, the 10% surface contribution on resonance accounts for the slightly less negative (less transverse) experimental angular contrast values in Fig. 2E relative to the purely volume Monte Carlo theory. With $\sigma_S^{(n\text{PPE})}/\sigma_V^{(n\text{PPE})}$ now determined and $\sigma_S^{(n\text{PPE})} + \sigma_V^{(n\text{PPE})}$ known directly from the total experimentally-measured photoemission rates (Eq. 2), we can determine $\sigma_S^{(n\text{PPE})}$ and $\sigma_V^{(n\text{PPE})}$ independently, as summarized in Fig. 4E for the representative nanorod. We note that $\sigma_S^{(n\text{PPE})}$ and $\sigma_V^{(n\text{PPE})}$ are only directly determined in the 3PPE range. While explicit $a^{(n)}$ values for other process orders could be determined via additional detuning studies, we instead simply rely on the approximate continuity in experimental $\sigma_S^{(n\text{PPE})}/\sigma_V^{(n\text{PPE})}$ values (i.e. no sudden changes observed with detuning) from the $L/D \approx 3$ nanorod detuning studies in Fig. 4 to extend the 3-photon results into the adjacent 2- and 4-photon ranges. With the aid of finite element and Monte Carlo modelling of the geometry-specific quantities (field integrals and escape efficiencies, respectively), this allows us in principle to now quantitatively predict the photoemission behaviors for arbitrary gold nanoparticle geometries.

## Discussion

We have shown that the plasmonic field enhancements are of primary importance in determining the relative surface and volume contributions to electron emission. With $a^{(3)}$ and therefore the material-specific properties in Eqs. 3A and 3B determined from detuning experiments, we can now quantitatively estimate the effects of optical parameters and nanoparticle geometry on the relative contributions of surface and volume photoemission, including the effects of (i) material screening, (ii) geometric surface-to-volume ($S/V$) ratio, and (iii) nanoparticle shape/curvature. In particular, we demonstrate that optical parameters influence the relative surface versus volume contributions primarily via frequency-dependent screening, while the nanoparticle shape controls the plasmonic field distribution at constant excitation frequency. Direct results from the 3PPE range (Fig. 4) are again extended into the 2PPE and 4PPE ranges by approximate continuity in $\sigma_S^{(nPPE)}/\sigma_V^{(nPPE)}$, although this will negligibly affect the principles and conclusions discussed.

First, we consider the effects of frequency-dependent screening and the plasmon resonance (Fig. 5A) on the surface vs. volume photoemission contributions. Specifically, we simulate the plasmonic field enhancements and resulting $\sigma_S^{(nPPE)}/\sigma_V^{(nPPE)}$ ratios for ligand-free nanorods in vacuum, for a series of aspect ratios (Fig. 5A). In this purely vacuum environment, the surface field enhancements are more prominent and longer nanorods are predicted to be surface emitters on resonance, unlike the ligand-coated, ITO-supported nanorods studied experimentally. For nanospheres in the electrostatic approximation and $\boldsymbol{E_0} = E_0\hat{x}$, one can analytically solve the Laplace equation to arrive at a constant internal field (50)

$$\boldsymbol{E_{\text{in}}}(\omega) = \frac{3\epsilon_0}{\epsilon(\omega) + 2\epsilon_0} E_0 \hat{x},$$

and a dipolar external field at the nanosphere surface,

$$\boldsymbol{E_{\text{out}}}(\omega) = E_0\hat{x} + \frac{\epsilon(\omega) - \epsilon_0}{\epsilon(\omega) + 2\epsilon_0}(3E_0\hat{n} - E_0\hat{x}),$$

where $\epsilon(\omega)$ is the dielectric function of the nanosphere material (e.g. gold) and $\hat{n}$ is the outward surface normal). The external field at the nanosphere "tip" along the laser polarization axis (i.e. $\hat{n} = \hat{x}$) is

$$\boldsymbol{E}_{\text{tip}}(\omega) = \frac{3\epsilon(\omega)}{\epsilon(\omega) + 2\epsilon_0} E_0 \hat{x},$$

from which the resulting ratio of the external-to-internal electric field at the tip is given simply by

$$\left|\frac{E_{\text{tip}}(\omega)}{E_{\text{in}}(\omega)}\right| = \left|\frac{\epsilon(\omega)}{\epsilon_0}\right|. \qquad [4]$$

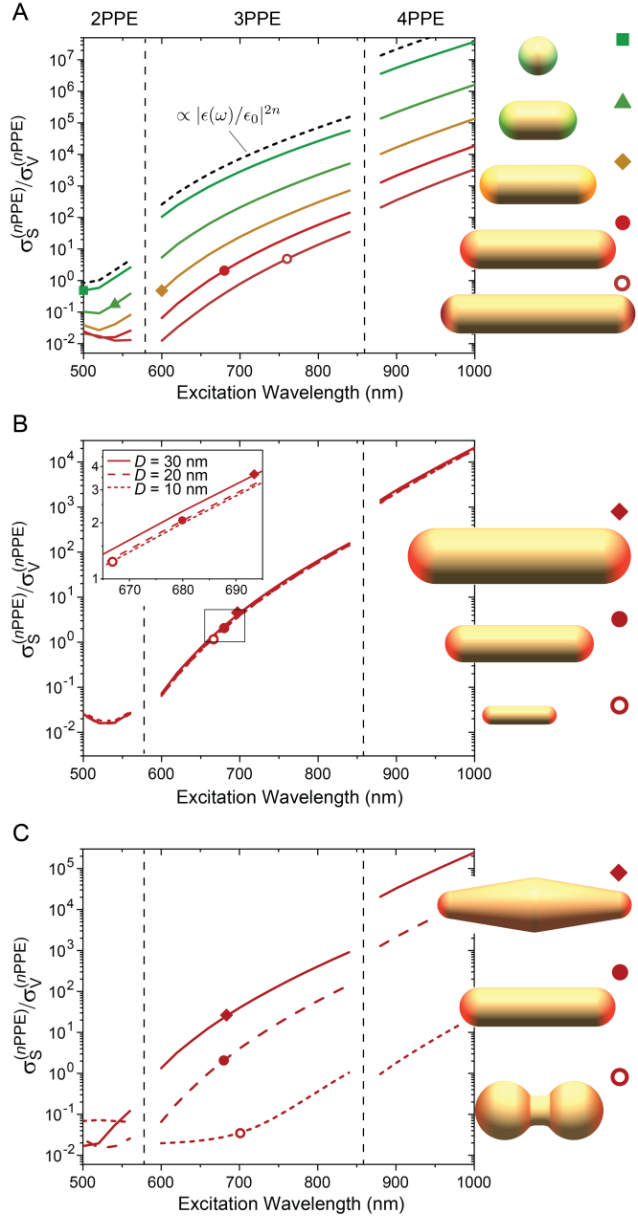

**Fig. 5.** Effects of screening, $S/V$ ratio, and nanoparticle geometry (curvature) on surface and volume contributions. This set of simulations is performed in vacuum (no ligands or substrate). (*A*) Surface/volume $n$PPE cross-section ratio determined quantitively from detuning studies (Fig. 4) for $L/D = 1\text{-}5$ nanorods. Points indicate plasmon resonance wavelength. The dashed line is the scaled dielectric function of gold, which determines the frequency dependence of the surface/volume cross-section ratios. (*B*) Cross-section ratio for $L/D = 4$ nanorod for $D = 10\text{-}30$ nm, with zoomed-in inset. (*C*) Cross-section ratio for bipyramid, nanorod, and dumbbell, all with similar resonance, $S/V$ ratio (to within 10%), and total $V$ (to within 30%).

This could also be immediately derived from the boundary condition on the displacement field, $|\boldsymbol{D}_{\text{in}} \cdot \hat{n}| = |\boldsymbol{D}_{\text{tip}} \cdot \hat{n}|$, yielding $|\epsilon(\omega)E_{\text{in}}| = |\epsilon_0 E_{\text{tip}}|$. The crucial result is that the plasmonic resonance term, $\propto (\epsilon(\omega) + 2\epsilon_0)^{-1}$, drops out of the field ratio in Eq. 4 entirely and we are left only with the screening effects described by the metal dielectric function, $\epsilon(\omega)$. Furthermore, the same cancellation of the surface and volume plasmonic field enhancements occurs in the full nonlinear field integrals, leading to $\sigma_S^{(n\text{PPE})}/\sigma_V^{(n\text{PPE})} \propto |\epsilon(\omega)/\epsilon_0|^{2n}$, as shown in Fig. 5A. We thus conclude that the dramatic increase in surface over volume photoemission at longer wavelengths can be simply attributed to enhanced metallic screening of internal fields. At a constant excitation wavelength, the progressive overall drop in $\sigma_S^{(n\text{PPE})}/\sigma_V^{(n\text{PPE})}$ with increasing nanorod aspect ratio in Fig. 5A is due to details of the nanorod geometry and plasmonic field distributions.

Next, we consider the effects of nanoparticle scale (Fig. 5B) on the surface vs. volume photoemission contributions. The importance of the geometric $S/V$ ratio has been emphasized in other investigations (13), but it remained unclear whether this was truly the decisive factor, in general, due to the previous lack of direct mechanistic insight into the surface and volume excitation processes. Here, we effectively isolate the effect of $S/V$ ratio by investigating nanorods with the same aspect ratio but different diameters, as in Fig. 5B. By maintaining the same shape, the field enhancement distributions and SPRs remain approximately constant with size, while the $S/V$ ratio changes by three-fold, from 0.44 nm$^{-1}$ ($D = 10$ nm) to 0.15 nm$^{-1}$ ($D = 30$ nm). This factor of three is already relatively minor compared to the many orders-of-magnitude changes in $\sigma_S^{(n\text{PPE})}/\sigma_V^{(n\text{PPE})}$ with modest detuning and the similarly strong shape-dependent effects to be discussed next. Moreover, a commensurate change in the volume emission efficiency with nanoparticle size, $1/\eta_V^{(n)} \propto D$, effectively cancels the $S/V \propto D^{-1}$ contribution to $\sigma_S^{(n\text{PPE})}/\sigma_V^{(n\text{PPE})}$. This occurs for larger particles ($D \gg \lambda_{\text{inel}}$) as the fraction of hot electrons excited within the escape depth from the surface ($\sim\lambda_{\text{inel}}$) to the total number of electrons excited throughout the nanoparticle approaches the geometric $S/V$ ratio (12), such that the escape efficiency scales as $S/V$ (i.e. $\eta_V^{(n)} \propto D^{-1}$). Thus, $\sigma_S^{(n\text{PPE})}/\sigma_V^{(n\text{PPE})}$ remains remarkably constant with nanoparticle scale at fixed excitation frequency, as shown in Fig. 5B. It is not until particle dimensions become comparable to or smaller than the hot electron inelastic mean free path that the $S/V \propto D^{-1}$ scaling begins to take over (12). This $\sigma_S^{(n\text{PPE})}/\sigma_V^{(n\text{PPE})} \propto D^{-1}$ regime may nonetheless be relevant for low-energy hot electron catalysis (e.g. $\lambda_{\text{inel}} \approx 40$ nm for gold at ~1.5 eV excitation energy) with small nanoparticles ($D < 30$ nm).

Finally, we consider the effects of nanoparticle shape (Fig. 5C) on the surface vs. volume photoemission contributions. While the effect of nanoparticle size/scale has been shown to be generally minor, the nanoparticle shape strongly influences $\sigma_S^{(n\text{PPE})}/\sigma_V^{(n\text{PPE})}$ via the plasmonic field distributions. Sharp convex features, for instance, can lead to more dramatic surface field enhancements due to the lightning rod effect (51, 52), whereas flat or concave features can shift emphasis to the volume fields. Thus, we attempt to isolate the effects of nanoparticle curvature by comparing particles of different shapes but similar SPR, $S/V$ ratio, and total volume (see Fig. 5C). Sharper-tipped geometries, such as bipyramids, have more concentrated surface field enhancements and more diffuse volume enhancements (*SI Appendix*, Fig. S13), leading to an enhanced surface photoemission contribution. This is corroborated by observations of tip-localized photoemission from silver bipyramids (53), and gold nanostars (8, 54), and etched gold nanotips (52). Conversely, more concave geometries such as dumbbells display much weaker surface fields and stronger relative interior field enhancements (*SI Appendix*, Fig. S13), leading to a dramatically enhanced volume photoemission contribution.

Interestingly, we note as a parting comment that the effect of shape on internal quantum efficiency, $\eta_V^{(n)}$, is typically negligible. For example, $\eta_V^{(3)} = 0.27\%$ for the $L/D = 4$ resonantly-excited nanorod in Fig. 5C, while $\eta_V^{(3)} = 0.32\%$ for the dumbbell excited at nearly the same SPR frequency. The strong influence of nanoparticle shape on $\sigma_S^{(n\text{PPE})}/\sigma_V^{(n\text{PPE})}$ therefore arises from the shape-dependent distribution and concentration of plasmonic fields, rather than the volume escape efficiency. This has two significant benefits for designer plasmonic devices: (i) the plasmonic fields can be readily simulated by a variety of classical methods (e.g. finite element simulation), and (ii) a further degree of optical control – beyond frequency-dependent screening – can be readily exerted by coupling to different plasmon resonance modes via laser polarization and frequency. Different plasmon resonance modes will display different volume and surface field distributions, which has been utilized, for instance, to control photocurrents from gold nanostars with multiple tip hot spots associated with different plasmon modes (8, 54).

## Conclusions

In summary, we have demonstrated the essential role of nanoscopic volume vs. surface photoexcitation mechanisms in nanoplasmonic hot electron emission, along with corresponding opportunities for designing and optically controlling hot electron spatial and momentum distributions. Volume excitation, which is dominant for nanorods excited at their longitudinal resonances, leads to hot electrons excited predominantly near the center of the nanorods in the centralized field-enhanced region, which subsequently escape from the nearby side surfaces in a transverse momentum distribution. However, red detuning of the excitation frequency strongly de-emphasizes the volume fields due to enhanced metallic screening, instead promoting hot electron excitation directly at the tip surfaces. The surface-excited electrons are preferentially emitted longitudinally along the nanorod axis and therefore exhibit completely different spatial and momentum distributions compared with the volume-excited hot electrons. Both processes are shown to be important in nanoplasmonic systems and must be accounted for in general. We have demonstrated that comprehensive volume (ballistic Monte Carlo) and surface (fully quantum) MPPE theory can be used to model these behaviors, but more importantly that the plasmonic field enhancement distributions (rather than the geometric $S/V$ ratio or internal quantum efficiency, $\eta_V^{(n)}$) are critical in controlling the surface vs. volume excitation. After characterizing the material-specific (nonlinear) absorbances, we have shown that the surface vs. volume photoemission properties of arbitrary gold geometries can be predicted simply via classical electrodynamics simulations. This introduces exciting opportunities for the design of hot electron catalysis and nanoelectronic devices, in which the geometry can be optimized to control plasmon mode structure and corresponding surface vs. volume distributions. Hot electron spatiotemporal distributions can then be controlled on nanometer spatial scales and femtosecond timescales via ultrafast optical frequency and polarization manipulation (2, 8, 55).

## Methods

### Nanorod Synthesis and Sample Preparation

Gold nanospheres and nanorods of various sizes and aspect ratios have been synthesized using well-known seed-mediated methods (56-58). Small gold precursor particles are prepared in the desired solvent (water, in this case) and exposed to additional gold in the presence of a reducing agent that promotes controlled growth onto the precursor particles. For nanospheres, the seed concentration in the growth solution is controlled for size uniformity, where more seeds result in smaller nanospheres. The gold nanorods are controlled in their size and aspect ratio by controlling both the seed concentration and the concentration of silver nitrate – a growth-directing agent – in the solution. The details of each synthesis procedure and the reagents used are described in detail in the *SI Appendix*, Section 1.

Following synthesis, the nanosphere and nanorod solutions are sonicated for 30 s and vortexed for 10 s for optimal dispersion, then diluted in ultrapure water to approximately 0.1 nM ($6 \times 10^{10}$ mL$^{-1}$). Immediately following dilution, a 50 μL aliquot is spin-coated onto a freshly UV-ozone cleaned ITO-coated coverslip (10 nm ITO sputtered on 170 μm borosilicate; Thin Film Devices, Inc.) at 1500 rpm for 5 minutes. For aqueous nanoparticle solutions, this procedure yields a typical coverage of 0.05 nanoparticles/μm$^{-2}$, or 20 particles in a $20 \times 20$ μm$^2$ scan area on average. Such coverages are ideal for efficient particle location and characterization via SPIM scans, while ensuring negligible probability of two particles overlapping within the same diffraction-limited excitation region. For particle location in correlated SPIM-SEM studies, a 30 nm Au alphanumeric grid (LF-400) is deposited onto all ITO/glass substrates via negative photomask lithography, prior to sample preparation.

**Scanning Photoelectron Imaging Microscopy (SPIM)**

Photoemission microscopy using a home-built scanning sample stage allows for spatially-resolved (diffraction-limited) photoemission mapping (4). Three quartered piezoelectric posts with capacitive sensor feedback provide fine scanning over a $30 \times 30$ μm$^2$ area, with $xyz$ piezo motors providing extended positioning over a larger (few-millimeter) range. Single particles are identified and studied individually, with the total photoemission rate measured as a function of laser frequency, intensity, and polarization. Combined with correlated SEM imaging and finite element simulation, this already provides detailed near-field information on the photoemission properties of the plasmonic nanoparticles. To get additional information on hot electron spatial and angular distributions, essential for the present studies, we employ a three-electrode velocity map imaging (VMI) electrostatic lens for transverse ($v_x, v_y$) photoelectron velocity mapping. The VMI lens provides a linear velocity-to-position mapping onto a spatially-resolved microchannel plate detector (two-plate chevron configuration), in which a single electron is multiplied up to $\sim 10^7$ electrons. The amplified electron signal is then accelerated onto a P47 phosphor screen and the fluorescence is imaged via CCD camera (1.3 megapixels, 20 FPS). Single event ($x, y$) coordinates are determined via centroiding and converted to ($v_x, v_y$) velocity with a calibration factor of 4150 m s$^{-1}$px$^{-1}$ measured previously (59). Further details on the SPIM system and velocity mapping can be found in previous work (59).

Femtosecond pulses are generated via a 75 MHz Ti:sapphire oscillator (KMLabs Swift, 675-1000 nm), with second harmonic generation (350-500 nm) and an optical parametric oscillator (KMLabs, 510-780 nm signal tuning range) providing broad tunability from the UV to the near-IR. A high-vacuum-compatible ($< 5 \times 10^{-7}$ Torr) reflective Cassegrain microscope objective (NA = 0.65) focuses the pulsed laser beams to a diffraction-limited spot (~500 nm spot diameter) on the sample at normal incidence, as shown in Fig. 1A. For optimal frequency tunability without spatial walk-off, no external prism dispersion compensation is utilized for the majority of the present studies. Group velocity dispersion in the system thus results in 100-200 fs pulse durations at the sample across the laser tuning range. As a check, several measurements have also been performed with dispersion-compensated (~50 fs) pulses and found to yield indistinguishable nanorod photoemission velocity distributions. Nanorods are cleaned prior to all studies via brief (~1 s) exposure to ~1 GW cm$^{-2}$ of second harmonic light (400 nm), which removes adlayers (i.e. water) that develop during brief sample exposure to ambient air (55).

**Scanning Electron Microscopy (SEM)**

Correlated SEM (FEI Nova NanoSEM 630, 10 kV, $< 1 \times 10^{-5}$ Torr, through-lens detector, field immersion mode) is performed on every gold nanorod and nanosphere investigated. The conductive ITO film on the glass substrate provides a route for charge dissipation during imaging. While no particle morphology changes are observed during imaging, micrographs are always collected following SPIM studies to avoid particle degradation due to electron beam exposure and concomitant amorphous carbon buildup. The Au reference grid is utilized to locate the nanoparticles in both SPIM and SEM, whereby the relative particle orientation is determined to within a few degrees by the grid orientation and the particle distribution within a grid area.

**Finite Element Method**

Finite element simulations of the plasmonic electric field enhancements are performed using the RF module in *COMSOL Multiphysics 5.4*. For supported nanorod and nanosphere simulations, a 1.5 nm CTAB (cetyltrimethylammonium bromide) ligand layer surrounding the particles is included in the simulations, along with an additional 0.5 nm gap between the ligand layer and the substrate to avoid extra-narrow domain regions. Overall, the rectangular domain consists of the glass substrate with a 10 nm ITO film, the vacuum superstrate, the gold nanoparticle with ligand layer, and a perfectly-matched layer surrounding the domain to prevent field reflection at the domain boundaries. Nanorods are modelled as perfect cylinders with hemispherical end caps, with diameter $D$ and total tip-to-tip length $L$. Triangular surface and tetrahedral volume meshing were constructed with near-uniform element size across the nanoparticle surface/volume, with the element side length (2 nm) selected to be much smaller than the nanoparticle dimensions and any electric field variation on the surface or within the volume. This is particularly essential, as the same mesh and calculated field values at the vertices are subsequently utilized for both volume and surface photoemission modelling. The optical constants of gold are taken from the literature (60) and determined for the ITO film via ellipsometry . We utilized $n_{\text{lig}} = 1.5$ and $k_{\text{lig}} = 0.25$ for the CTAB ligand layer, where the small extinction coefficient accounts for the presence of amorphous carbon due to the hot-electron-driven conversion of the organic ligands. Further discussion of this conversion process and nanorod photoemission stability can be found in the *SI Appendix*, Section 7.

**Code and Materials Availability**

All data required to reproduce the results presented here are available within the paper, the *SI Appendix*, or from the corresponding authors upon reasonable request. All code is available from D.J.N. upon reasonable request.

**Associated Content**

This article contains supporting information.

**Author Information**

**Author Contributions**

J.P., F.M., and D.J.N designed the experiments. J.P. executed the experiments, performed the analysis, and developed the theoretical methods under the supervision of D.J.N. S.M.M. synthesized and characterized the nanorods and nanospheres under the supervision of C.J.M. J.P. wrote the manuscript with contributions and final approval from all authors.

**ORCID iDs**

Jacob Pettine: 0000-0003-2102-1743
Sean M. Meyer: 0000-0003-0771-5095
Fabio Medeghini: 0000-0002-0508-3365
Catherine J. Murphy: 0000-0001-7066-5575
David J. Nesbitt: 0000-0001-5365-1120

**Notes**

The authors declare no competing interests.

## Acknowledgements

This work has been supported by the Air Force Office of Scientific Research (FA9550-15-1-0090) with additional funds for laser and apparatus development provided by the National Science Foundation (Physics Frontier Center Program, PHY-1734006). Support for nanoparticle synthesis and characterization in the C.J.M. laboratory has been funded by the National Science Foundation (CHE-1608743). The authors would like to acknowledge Dr. J.G. Hinman's contributions to the synthetic procedures and discussions regarding this work.

# Supporting Information for

# Controlling Hot Electron Spatial and Momentum Distributions in Nanoplasmonic Systems: Volume and Surface Effects

Jacob Pettine[1,2], Sean M. Meyer[4], Fabio Medeghini[1], Catherine J. Murphy[4,5], and David J. Nesbitt[1,2,3]

[1]JILA, University of Colorado Boulder and National Institute of Standards and Technology, Boulder, Colorado 80309, United States

[2]Department of Physics, University of Colorado Boulder, Boulder, Colorado 80309, United States

[3]Department of Chemistry, University of Colorado Boulder, Boulder, Colorado 80309, United States

[4]Department of Chemistry, University of Illinois at Urbana-Champaign, Urbana, Illinois 61801, United States

[5]Materials Research Laboratory, University of Illinois at Urbana-Champaign, Urbana, Illinois 61801, United States

# 1) Nanorod Synthesis and Characterization

Nanospheres of 70 nm diameter are synthesized based on the method of Perrault *et al.* (1). A solution of 12 nm spherical gold seeds is prepared by quickly injecting 1 wt% sodium citrate to a boiling aqueous solution of 0.5 mM HAuCl₄ under vigorous stirring. After 5 minutes, a ruby red color is seen, and the heat is turned off and the solution allowed to cool naturally to room temperature on the hot plate. Without any purification, 9.6 mL of water is mixed with 100 µL of 1 wt% HAuCl₄ in a 20 mL glass vial and kept under continuous stirring at room temperature with a Teflon-coated stir bar. Quickly, 25 µL of 1 wt% sodium citrate and 115 µL of seed solution were added followed by a rapid injection of 100 µL of 0.03 M hydroquinone. After 20 minutes of stirring, the stir bar is removed, and the particles are centrifuged once at 1000 rcf for 15 minutes and dispersed into 1 mL water. This 1 mL of particle solution is quickly injected into 9 mL of 10 mM CTAB (cetyltrimethylammonium bromide) and mixed gently overnight. The following morning, the particles are subjected to two rounds of centrifugation at 1000 rcf for 15 minutes followed by redispersion in 5 mL of 1 mM CTAB.

Nanorods with aspect ratios ($L/D$) from 1.5-3 are synthesized using the seed-mediated growth method of Liz-Marzan and coworkers (2). Small nanorod seeds are prepared in high quality and used as the seeds to grow larger, monodisperse nanorods with the correct size and width. It is recommended, due to the length of the procedure, to refer to the original article for a complete understanding of the synthesis. Briefly, a CTAB and 1-decanol mixture is made and used to prepare small spherical seeds, then carefully in another binary surfactant solution they are introduced with more gold, silver, and acid to grow small nanorods for use in the growth of larger particles. Then, another similar growth solution containing silver, gold, hydrochloric acid, weak reducing agent, and seeds at a precise concentration is mixed and allowed to grow overnight. For the $L/D = 1.5$ nanorods 45 µL 0.01 M AgNO₃ and 55 µL of seeds are used with 500 µL 0.1 M hydroquinone in the absence of any additional acid. For the $L/D = 2.5$ nanorods 200 µL of 0.01 M AgNO₃, 50 µL of 1 M HCl, 80 µL if 0.1 M ascorbic acid, and 55 µL seeds were added. The nanorods are then subjected to centrifugation two times at 2000 rcf for 20 minutes and dispersed in 1 mM CTAB after 3+ hours of growth at 27 °C.

Nanorods with $L/D = 3-5$ are prepared, with variations, from the method of Zubarev *et al* (3). The first step is to make small CTAB-capped gold seeds of 1-2 nm. This is achieved by adding 0.46 mL of a basic sodium borohydride solution (prepared by adding 46 mg of sodium borohydride directly to 10 mL ice-cold 0.01 M sodium hydroxide and diluting this solution 10-fold with 0.01 M sodium hydroxide) under vigorous stirring to a solution of 9.5 mL 0.1 M CTAB and 0.5 mL 0.01 M HAuCl$_4$. After allowing an hour for the excess sodium borohydride to decompose, growth solutions are prepared. Growth solutions are prepared by mixing the following reagents as listed in order of 9.5 mL of 0.1 M CTAB, 0.5 mL of 0.01 M HAuCl$_4$, varying amounts of 0.1M AgNO$_3$ (10-40 µL), 0.5 mL of 0.1 M hydroquinone, and a varying seed amount (90 µL seeds for $L/D = 3$-4.5 and 40 µL for larger aspect ratio). The particles are kept at room temperature overnight and grown to completion. The next day, each tube is then subjected to centrifugation two times at 2000 rcf for 20 minutes and the nanorods are dispersed in 1 mM CTAB after 3+ hours of growth at 27 °C.

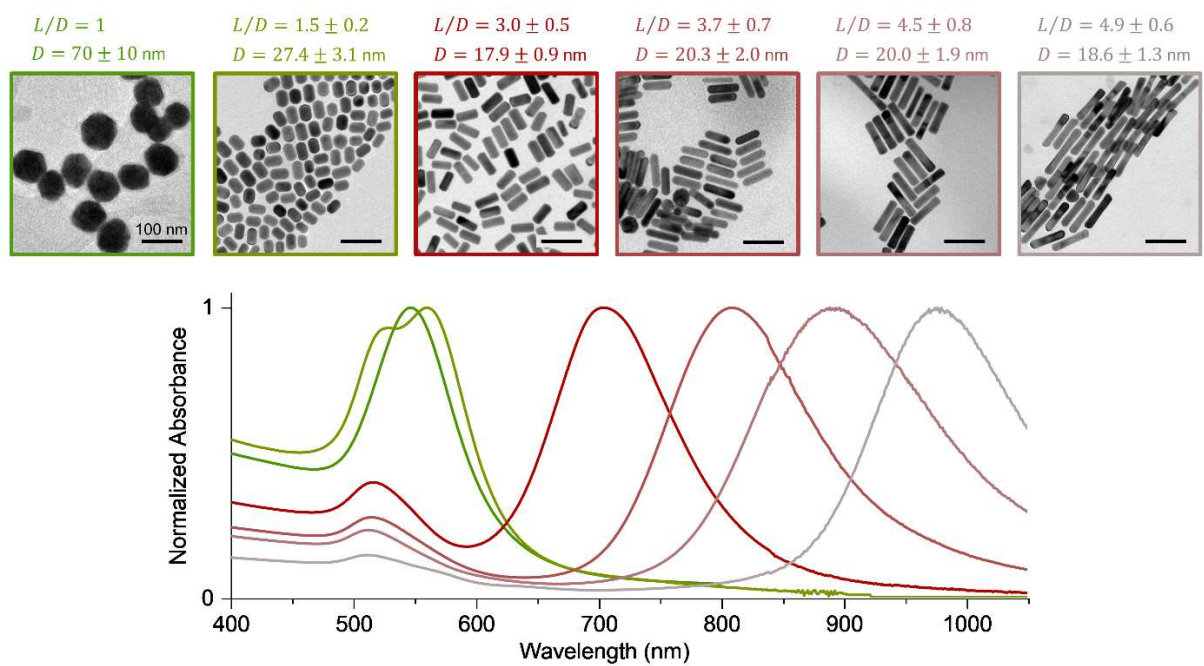

**Fig. S1.** Transmission electron micrographs and dimensional statistics of synthesized CTAB-coated nanorod samples, with aspect ratios ranging from 1 (spheres) to 5. The spheres are larger to ensure sufficient signal-to-background in photoemission measurements. Corresponding UV-Vis spectra for aqueous dispersions show longitudinal surface plasmon resonance peaks ranging from 550-980 nm, with transverse surface plasmon resonance peaks at 510 nm.

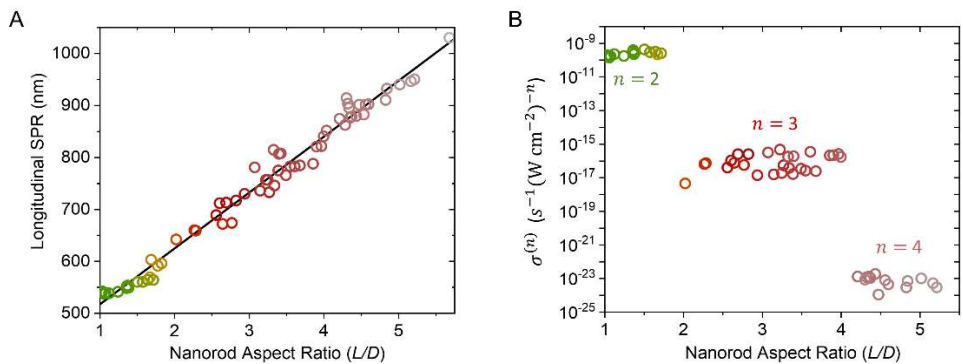

**Fig. S2.** Photoemission characteristics of the nanorod samples. (*A*) Surface plasmon resonances measured via multiphoton photoemission as a function of nanorod aspect ratio. The black line is a linear least squares fit. (*B*) *n*-photon photoemission cross-sections determined as a function of nanorod aspect ratio via intensity-dependence studies.

## 2) Multiphoton versus Thermionic or Optical Field Emission

The predominant role of multiphoton photoemission (MPPE) in the excitation/energy ranges utilized in the present studies is evidenced in two ways. First, power-law intensity-dependence behavior (Fig. S3) always has a clear integer power in the expected range given energy conservation, where

$$\Gamma_{\text{MPPE}} = \sum_n \Gamma_{n\text{PP}} = \sum_n \sigma_n(\omega, \theta) I_0^n,$$

and log-log intensity-dependence slope

$$\frac{d\log_{10}(\Gamma_{\text{MPPE}})}{d\log_{10}(I_0)} = \frac{\sum_n \sigma_n(\omega, \theta) I_0^n n}{\sum_n \sigma_n(\omega, \theta) I_0^n} = \sum_n w_n n.$$

Non-integer powers in the transition regions therefore represent weighted average over process orders with weight factors (relative process order contributions) $w_n = \sigma_n(\omega, \theta) I_0^n / \sum_{\tilde{n}} \sigma_{\tilde{n}}(\omega, \theta) I_0^{\tilde{n}}$, such that $\sum_n w_n = 1$. Second, the measured velocity map Fermi edges in good agreement with $n\hbar\omega - \phi$ (i.e. the $n$-photon excitation energy minus the work function). This remains true for the off-resonant detuning studies, even for the further red detuning (see Section 6).

Alternative possible emission mechanisms include thermionic emission from the transiently-heated electron gas or strong-field tunneling emission, known as optical field emission. A substantial benefit of working in the weak-field multiphoton intensity regime is avoiding excessive thermal effects. To definitively rule out thermionic emission or other heating effects in these studies, electron and phonon (lattice) temperatures are calculated for gold nanorods under femtosecond-pulsed excitation using the two-temperature model (4). Excited plasmons decay into electron-hole pairs, leading to heating of the electron gas (Supplementary Figure 6a), which thermalizes via electron-electron scattering on ~100 fs timescales before any substantial energy transfer to nanoparticle lattice occurs. Electron-lattice thermalization occurs on few-picosecond timescales via electron-phonon scattering (Fig. S4). Approximating the pulse energy to be directly transferred into the thermalized electron distribution, the coupled two-temperature equations are

$$C_e(T_e) \frac{dT_e}{dt} = -g(T_e - T_l) + I(t)\sigma_{\text{abs}},$$

$$C_l \frac{dT_l}{dt} = g(T_e - T_l),$$

with electron-phonon coupling constant, $g = 2 \times 10^{-7}$ W K$^{-1}$ (4, 5), Gaussian pulse intensity profile, $I(t)$, nanorod absorption cross-section, $\sigma_{\text{abs}}$, electron heat capacity, $C_e$, and lattice heat capacity, $C_l$. The temperature-dependent free-electron Sommerfeld heat capacity (6) is $C_e(T_e) = \pi^2 k_B^2 T_e n_e/(2E_F) = 1.5 \times 10^{-19}$ J K$^{-1}$, with free electron density $n_e = 5.9 \times 10^{28}$ m$^{-3}$ and Fermi energy $E_F = 5.53$ eV for gold. The lattice heat capacity, $C_l = c_l \rho V = 2.5 \times 10^{-17}$ J K$^{-1}$, is calculated using the specific heat of bulk gold, $c_l = 129$ J K$^{-1}$kg$^{-1}$, density of gold $\rho = 19.32$ kg m$^{-3}$, and nanoparticle volume $V$. The lattice heat capacity is over two orders of magnitude larger than the electron heat capacity, which explains the negligible peak lattice temperature (~325 K) compared with the peak electron temperature (~1500 K) calculated in Fig. S4 for a $D = 20$ nm, $L/D = 3$ nanorod. The thermionic surface current is calculated via the Richardson-Dushman equation (6)

$$J = AT_e^2 e^{-\frac{\phi}{k_B T_e}}$$

where $\phi$ is the work function and $A$ is the Richardson constant, $A = 7.5 \times 10^{24}$ electrons m$^{-2}$s$^{-1}$K$^{-2}$. Integrating the thermionic current over the nanorod surface during a pulse-stimulated electron heating cycle yields $5 \times 10^{-4}$ electrons s$^{-1}$, which is much smaller than typical multiphoton photoemission currents of 100-1000 electrons s$^{-1}$. Thus, thermionic emission is both calculated and observed to be negligible in the present studies.

Strong-field tunneling emission generally occurs for surface fields > 1 V Å$^{-1}$, where the electric field approaches atomic field strengths, as can occur due to plasmonic field enhancements. Optical field emission becomes dominant when the optical oscillation frequency is much smaller than the inverse electron "tunneling time", $\omega_t = eE/\sqrt{2m_e \phi}$, as characterized by the Keldysh parameter (7),

$$\gamma = \omega/\omega_t = \sqrt{\phi/2U_p},$$

for ponderomotive energy $U_p = e^2 E^2/(4m_e \omega^2)$ with electric field $E$. Perturbative MPPE is dominant for $\gamma > 2$ and optical field emission is dominant for $\gamma < 1$, with the transition occurring in the $1 < \gamma < 2$ range (8). For peak input pulse intensities $I_0 < 1 \times 10^8$ W cm$^{-2}$, $\phi \approx 4.5$ eV, and simulated field enhancements $|E/E_0| < 25$ for resonantly-excited nanorods, Keldysh parameters $\gamma > 30$ fall well within the MPPE regime in the present studies.

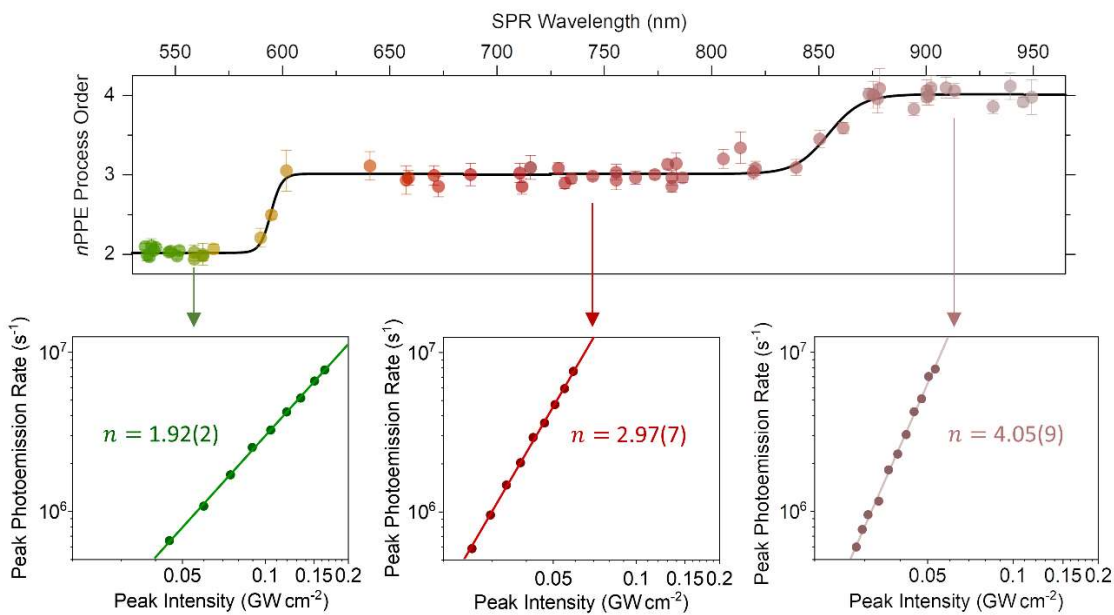

**Fig. S3.** Intensity-dependence process order summary with example single-nanorod data and fits for $n$ = 2, 3, 4. All measurements are performed at the nanorod longitudinal surface plasmon resonance.

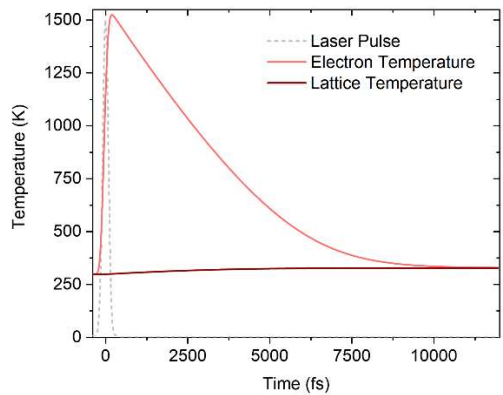

**Fig. S4.** Two-temperature model of electron and lattice heating for a $D = 20$ nm, $L/D = 3$ nanorod. The peak lattice temperature increase is ~30 K, while the conduction electrons have a pulse-averaged temperature of ~1000 K (at the pulse peak) and reach an overall peak temperature of 1500 K a couple hundred femtoseconds following excitation, before thermalizing with the lattice on a few-picosecond timescale.

S7

## 3) Effects of Nanorod Diameter and Surface Ligands

Nanorods of various diameter ($D$ = 10-40 nm) have been studied and all display similar transverse emission at longitudinal resonant excitation (Fig. S4). Nanorods of different sizes can have very different faceting (9). Furthermore, nanorods of two different ligand coatings – cetyltrimethylammonium bromide (CTAB) and hexadecanethiol (HDT) – were studied and also display similar emission behaviors. We therefore conclude that surface properties, such as facet-specific work function or scattering, do not strongly influence the fundamental volume-dominant excitation/emission behaviors demonstrated here at resonant excitation.

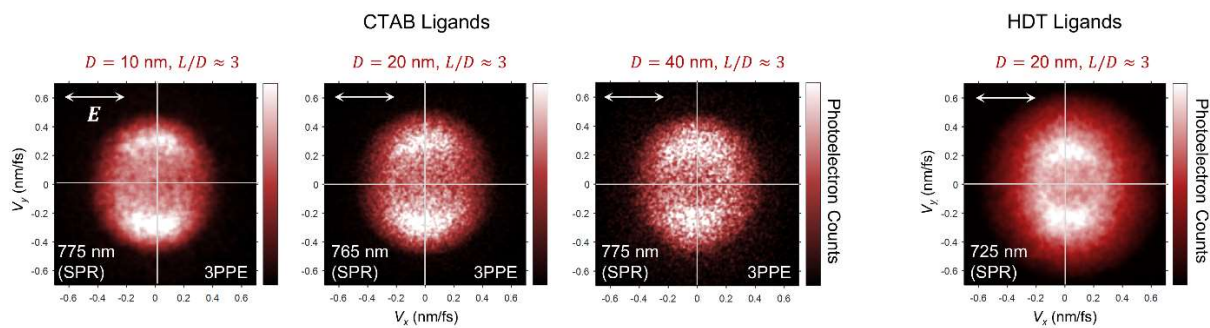

**Fig. S5.** Velocity map images collected for CTAB-coated nanorods of 10 nm, 20 nm, 40 nm diameters with similar aspect ratios and longitudinal surface plasmon resonance frequencies. The rightmost panel shows the velocity map for a 20 nm diameter nanorod coated in HDT ligands, which have a covalent thiol-gold bond rather than an ionic CTA-bromide bond.

# 4) Volume Photoemission Modelling

Volume (multiphoton) photoemission is modelled within the phenomenological, ballistic three-step framework developed by Berglund and Spicer in 1964 (10). Although phenomenological, the three-step model was placed on firm theoretical footing by Feibelman and Eastman (11), and others (12), via comparison with a formal one-step Fermi's Golden Rule treatment. In the three-step framework, electrons are (i) optically excited into a higher-energy eigenstate of the material, (ii) travel ballistically to the surface with an exponential survival probability to inelastic scattering ($\lambda_{\text{inel}} \approx 5$ nm at $E_F + 5$ eV in gold (13, 14)) along the way, and (iii) transmitted into the vacuum (or other surrounding collection medium, such as semiconductor or surface adsorbate layer) with finite probability if they have sufficient normal momentum to overcome the surface potential barrier. For a step potential barrier, the transmission function increases gradually above threshold and is given by (15)

$$T_{\text{step}}(k_z) = \frac{4\hbar k_z p_z}{\hbar k_z + p_z},$$

where $\hbar k_z$ is the internal surface-normal momentum and $p_z$ is the external surface-normal momentum, related via energy conservation by

$$\frac{\hbar^2 k_z^2}{2m_e} = \frac{p_z^2}{2m_e} + E_F + \phi,$$

for surface barrier height $E_F + \phi$. By contrast, the transmission function for a smooth barrier due, for instance, to an unscreened image force, is approximately a unit step function (15)

$$T_{\text{smooth}}(k_z) \approx \theta\left(\frac{\hbar^2 k_z^2}{2m_e} - E_F - \phi\right)\theta(k_z) = \theta(p_z).$$

We utilize the smooth-barrier transmission step function for all volume photoemission calculation in the present work.

To account for all hot electron trajectories over a nanoparticle volume within the three-step photoemission framework, we implement a Monte Carlo numerical integration routine. Whereas direct integration is computationally intensive, wasteful for highly spatially-nonuniform nonlinear excitation, and also requires care with the internal-to-external phase space Jacobian transformation, the Monte Carlo method is efficient and encodes all Jacobian effects automatically. Varying degrees of sophistication have been implemented in previous Monte Carlo hot electron transport/emission calculations, including emphasis on surface-scattering (16) or on volume scattering determined via *ab initio* theory (17, 18). None of these previous calculations have emphasized full momentum resolution, however, only total incident counts or internal quantum efficiencies. Here we implement the elements that are most important for full 3D velocity resolution and nonuniform spatial excitation (i.e. utilizing near-fields determined via finite element simulation) with inelastic mean free paths smaller than particle dimensions (thus neglecting minor surface scattering effects). We approximate a constant joint density of states and constant excitation matrix elements for the coherent nonlinear excitations, such that the ground state Fermi-Dirac distribution is preserved for the nascently-excited hot electrons. Furthermore, we approximate that the hot electrons are excited isotropically, which is realistic for phonon-mediated multiphoton excitation that effectively randomizes the final hot electron momentum.

The Monte Carlo routine is represented in Fig. S6. A weighted random selection of a nanorod volume point is performed with nonlinear field enhancement weight factor $|E/E_0|^{2n}$ determined via finite element simulation. To avoid volume discretization error, we then perform a random displacement ($\Delta r$) within the volume associated with each mesh vertex. An angle is then selected at random and the momentum is randomly selected, weighted by the excited Fermi-Dirac distribution. Next, the surface vertex closest in angle is determined and the corresponding distance and surface normal ($\hat{n}_{surf}$) are used to determine scattering and transmission probabilities, respectively. Given the small amount of excess kinetic energy (~1 eV) relative to the work function (~4.25 eV), hot electrons that undergo even a single inelastic scattering event with a cold electron and lose half of their energy on average (19) and are therefore typically unable to escape and may be neglected. We do account for the possibility of surviving a single scattering event via the triangular final energy distributions of electrons determined by Ritchie and Ashley (19), accounting for exchange effects, but find the emission contribution of these scattered electrons to be negligible. The mean free path for elastic scattering (~30 nm) is much larger than the inelastic mean free path (~5 nm) and therefore safely neglected as well. In cases where it is deemed necessary – i.e. for lower-energy excitations escaping over a low-energy Schottky barrier – the effects of inelastic and elastic scattering can be readily included in this simple Monte Carlo algorithm. This method can also be modified to allow for geometries with concave features (and therefore self-intersecting trajectories), but the simple convex modelling covers many essential geometries, including spheres, rods, cubes, triangles, etc.

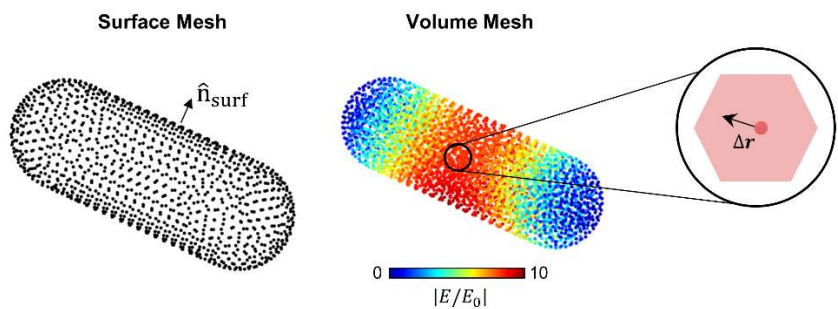

**Fig. S6.** Depiction of the surface and volume mesh vertices utilized in the Monte Carlo modelling. Each surface vertex corresponds to a ~1 nm² area with nearly constant surface normal ($\hat{n}_{surf}$). Each volume vertex corresponds to a ~1 nm³ volume with an electric field enhancement determined via finite element modelling. To avoid any discretization effects, we randomly perturb the excitation coordinates by $\Delta r$ within the volume associated with a randomly-selected (nonlinear field-weighted) mesh vertex.

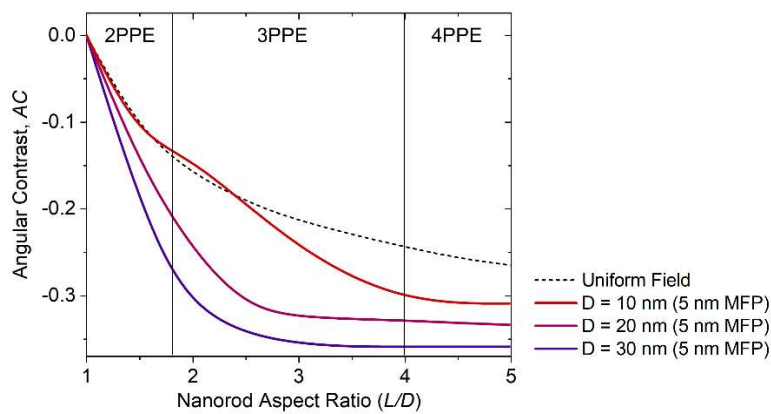

**Fig. S7.** Calculated angular contrast ($AC$) values for nanorods of different diameters and aspect ratios, using the nonlinear field distributions calculated via finite element modelling. As the nanorod radius becomes comparable to the inelastic mean free path (~5 nm here) more electrons can escape from the entire nanorod surface, including the hemispherical tips. This leads to a more isotropic emission distribution and less negative (less transverse) $AC$ values. The uniform field curve only depends on the relative side vs. tip surface areas, which depends only on the aspect ratio and is constant with diameter.

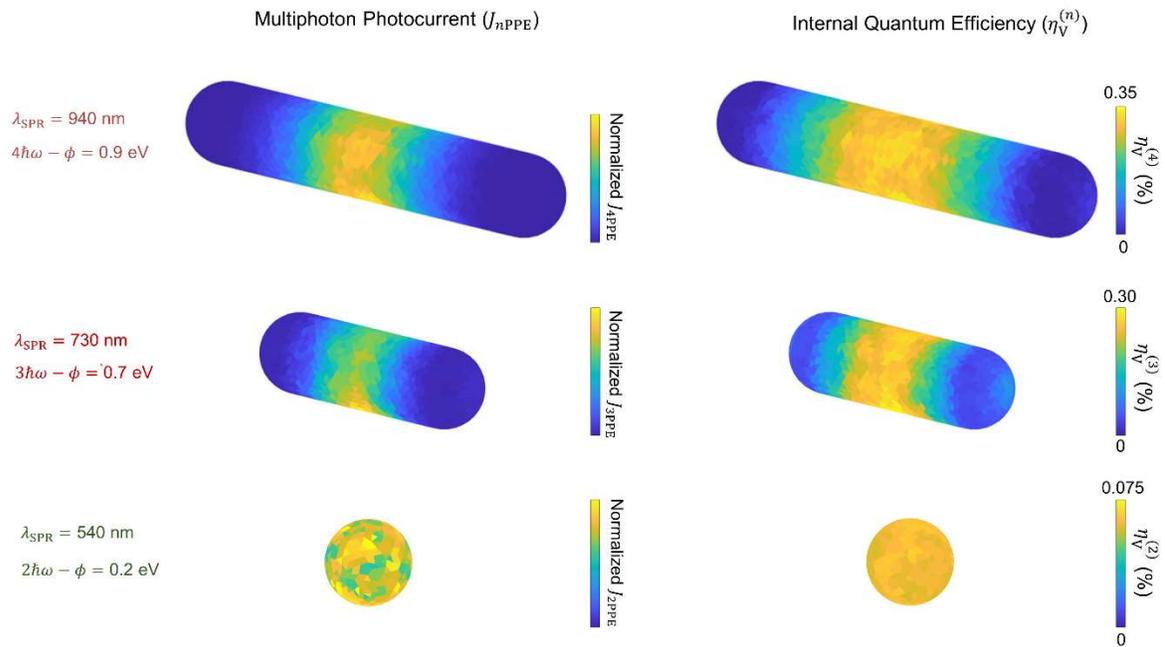

**Fig. S8.** Surface photocurrent ($J_{n\text{PPE}}$) and internal quantum efficiency maps for resonantly-excited nanorods with aspect ratio $L/D = 5$ (top), 3 (middle), and 1 (bottom), determined via Monte Carlo volume photoemission simulations. Each rod is resonant in a different nonlinear regime. (Left) The surface current proportional to the total number of escaped electrons from each surface area element, which shows that most electrons are excited near the center of the nanorod in the centralized field-enhanced region, subsequently escaping mostly from the nearby surfaces (as limited by inelastic scattering). (Right) The internal quantum efficiency is taken to be a discrete value for each area element, with the total given by surface-averaged value. The $L/D = 5$ and 3 rods have similar maximum electron kinetic energies ($n\hbar\omega - \phi$) of ~0.8 eV at resonance and therefore have similar quantum efficiencies. The excess electron kinetic energy is much smaller (~0.2 eV) in the sphere case for the 540 nm resonance and 2-photon photoemission, such that the escape quantum efficiency is about four times smaller. Such frequency dependence is described by Fowler theory (20) (with the frequency dependence trivially extended into the $n$-photon regime via $\hbar\omega \to n\hbar\omega$).

# 5) Surface Photoemission Modelling

We apply the metal surface multiphoton photoemission theory developed by Yalunin and coworkers (21) to the case of nanoparticle photoemission using the surface field distributions determined via finite element modelling. While this has been described in detail previously (22), we will reiterate some of the essential details here. The probability for excitation from the external evanescent tail of free conduction electron wavefunctions into external field-dressed Volkov states is solved perturbatively via Green's functions (21). Field penetration into the nanoparticle is neglected, such that all of the excitation density is external. The ponderomotive field dressing is accounted for in the energy conservation relation,

$$\frac{\hbar^2 k^2}{2m_e} + n\hbar\omega = \frac{p^2}{2m_e} + U_\text{p} + E_\text{F} + \phi,$$

where $k$ is the momentum of the ground state and $p$ is the external free electron momentum. However, the ponderomotive energy in the present case is negligible (< 0.1 eV) and is therefore neglected. We calculate the 3D photoemission distribution for each surface area element, then sum over the entire surface. The surface mesh is uniform and much finer than any variation in the plasmonic electric field.

## 6) Red Detuning Effects

The transition from volume photoemission (angular contrast, $AC < 0$) on/around the surface plasmon resonance to surface photoemission ($AC > 0$) with red detuning of the excitation laser is shown for nine $L/D \approx 3$ nanorods in Fig. S9. The resonances of the nanorods range from 715 nm to 800 nm (a ~0.15 eV spread). With respect to detuning from the surface plasmon resonance, all curves are nearly overlapped. However, the spread in the curves with respect to absolute excitation energy is as broad as the spread in resonances. Thus, it is clear that the transition cannot be due to an absolute energy effect such as the transition from 3PPE to 4PPE (with this line noted in the right panel of Fig. S9). We have also noted in the main text, however, that the plasmon resonance effect, $\propto (\epsilon(\omega) + 2\epsilon_0)^{-1}$, drops out of the expression for the surface-to-volume nonlinear cross-section ratio, $\sigma_S^{(nPPE)}/\sigma_V^{(nPPE)}$, which thus depends only on $|\epsilon(\omega)/\epsilon_0|^{2n}$. While this would appear to be an absolute frequency effect, the geometry itself changes the distribution such that $\sigma_S^{(nPPE)}/\sigma_V^{(nPP\ )}$ happens to be similar for all nanorods on resonance. This is merely a coincidence that is not true, for instance, for ellipses, or even for nanorods simulated in vacuum (as shown in the Discussion section of the main text). Thus, the categorization as a "detuning" effect versus an "absolute energy" effect comes with a bit of a caveat, but we primarily want to show that it is not an effect due to the process order transition. Thus, we should really refer to it as a "screening" effect.

At far red detuning from the plasmon resonance, the plasmonic field enhancement is much smaller and must therefore be compensated by larger input intensities to achieve similar photoemission signal rates (and thus collect data in a reasonable amount of time). The increased intensity therefore doesn't lead into the strong-field regime, which depends on the input field augmented by the plasmonic field enhancement. To verify that red-detuned photoemission is still perturbative (multiphoton) in nature, rather than strong-field or thermionic, we show the 4PPE intensity dependence in Fig. S10.

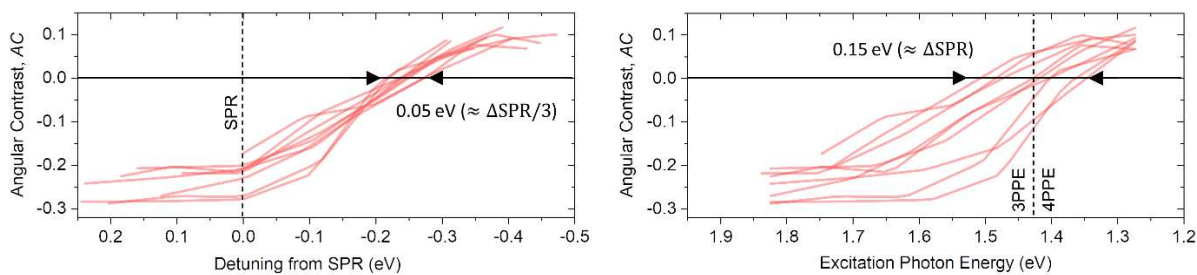

**Fig. S9.** Demonstrating that the transition from volume to surface photoemission at lower excitation frequencies is not due to an absolute energy effect, but rather a detuning effect. (Left) Angular contrast versus detuning from the surface plasmon resonance, as also shown in the main text. All curves for 9 nanorods have similar trends and overlap quite well. The spread in detuning values crossing into the surface regime $AC = 0$ is a factor of three smaller than the spread of surface plasmon resonance values. (Right) Angular contrast versus absolute excitation photon energy, showing a much broader spread dictated entirely by the spread in SPRs of the 9 nanorods.

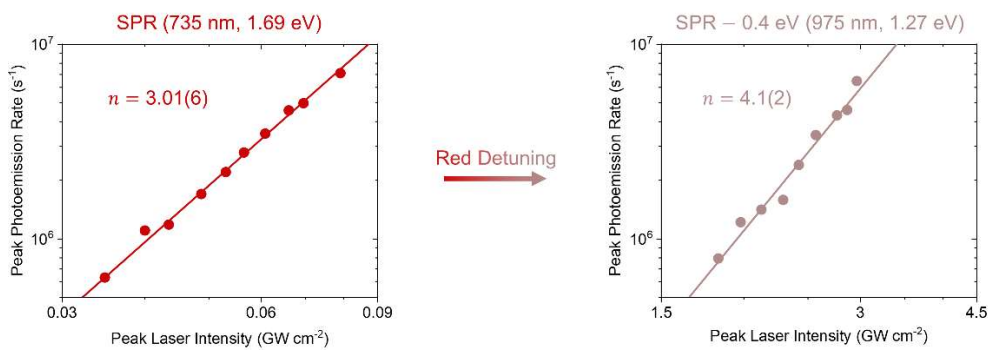

**Fig. S10.** (Left) 3PPE intensity dependence for an example nanorod ($L/D \approx 3$) excited on resonance, versus (Right) the 4PPE intensity dependence. Red-detuned emission therefore remains in the multiphoton regime, as expected.

# 7) Nanorod Stability

Under the present weak-field input intensities, nanorod heating is negligible (~30 K lattice temperature increase, Fig. S4) and thus there is no possibility of melting. To verify that nanorods remain stable emitters during these studies, we demonstrate three typical time traces in Fig. S11 for nanorods excited at their surface plasmon resonances (SPRs) and at red detuning. The volume emission on resonance is clearly very stable, while surface emission at red detuning can be quite spiky, if fundamentally still fairly stable. The cause of the spikes is unknown, though may be attributed to single-atom displacement (so-called "picocavity") effects, which transiently create a factor of ~4 larger field in a sub-nanometer region (23). For the 4PPE process, the 4-fold increase in $|E/E_0|$ leads to a $10^5$-fold increase in $|E/E_0|^8$, which can thus have noticeable effects despite the miniscule single-atom area. The surface signal fluctuations may alternatively be attributed to ligand rearrangement and corresponding sudden changes in the surface dielectric environment. In any case, the volume emission is insensitive to such effects and thus remains quite stable.

Looking with velocity resolution, we find that while volume photoemission distributions remain quite stable/unchanged after prolonged nanorod exposure, surface distributions tend to change a bit after ~1 hour of continuous exposure (Fig. S12). We attribute this to well-known buildup of amorphous carbon in the strongly field-enhanced tip regions (24), in this case attributed to the cracking and rearrangement of surface ligands. The nonuniform amorphous carbon leads to varying near-field enhancements and corresponding surface photoemission distributions, while potentially also increasing scattering and therefore leading to more diffuse distributions.

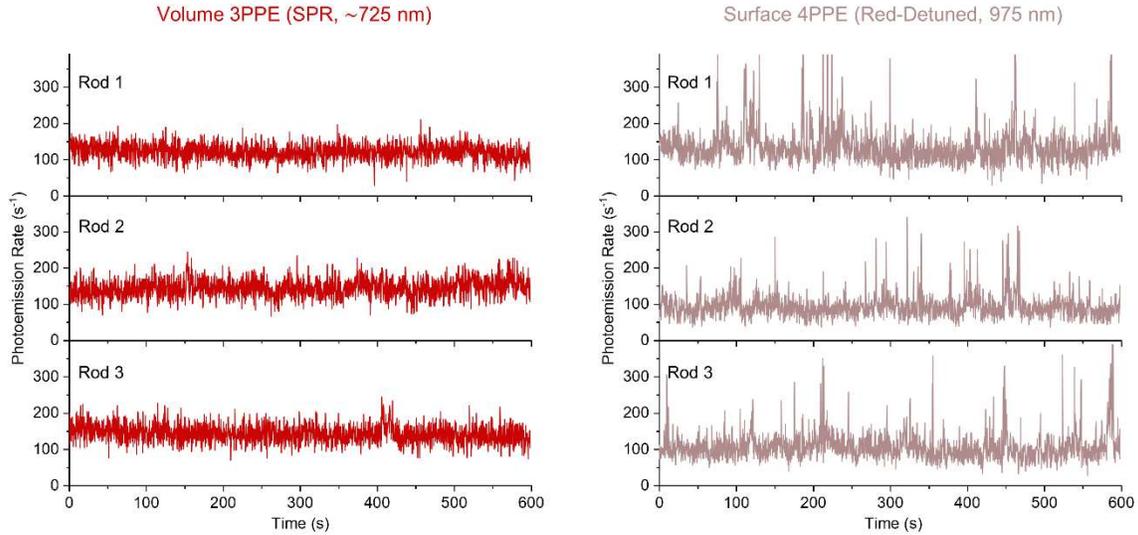

**Fig. S11.** Volume (left) and surface (right) MPPE time traces for three typical nanorods.

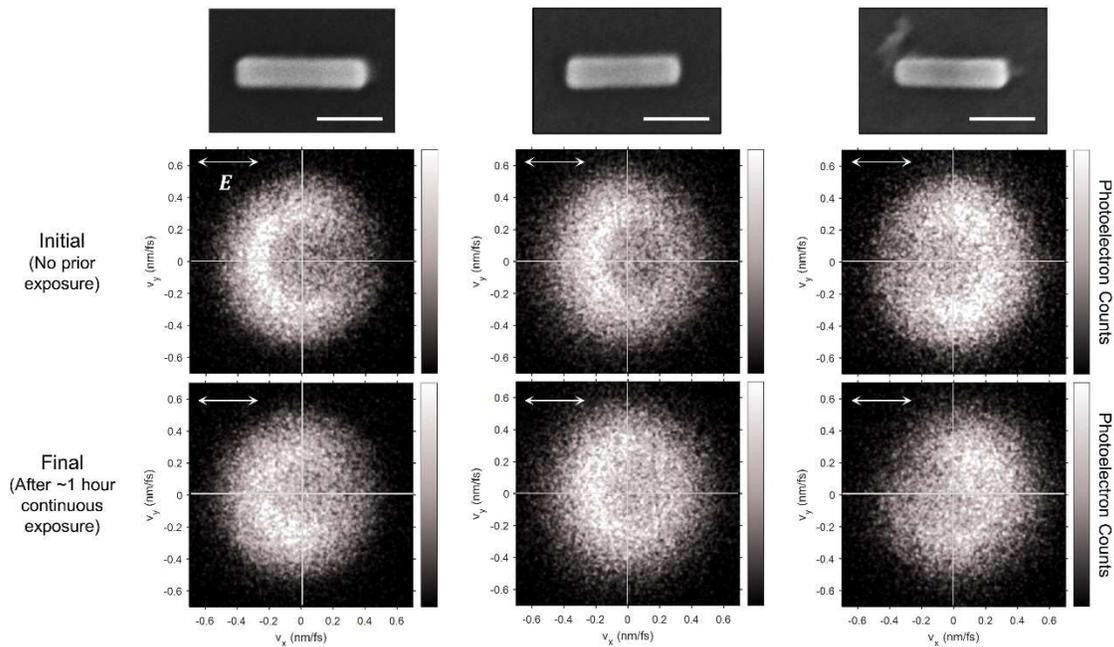

**Fig. S12.** Typical changes observed in the surface MPPE velocity distributions following ~1 hour of continuous laser exposure at typical intensities. All three $L/D \approx 4$ nanorods shown here are excited at 975 nm ($\approx$ SPR $- 0.2$ eV) in the 4PPE regime, with angular contrast $AC \approx 0$. Due to the limited detuning from resonance (~0.2 eV), the distributions still have a notable transverse component superposed with the asymmetric longitudinal surface emission.

# 8) Surface and Volume Field Distributions

Slices through the finite element simulation domain are shown in Fig. S13 for bipyramid, nanorod, and dumbbell geometries discussed in the main text. Not only does the surface field enhancement decrease when going from sharper (more convex/prolate) to flatter (or even concave) shapes, but the volume field is more concentrated and actually increases in the central region of the dumbbells. This leads to a strong preference for surface emission from the bipyramids, while the dumbbells are strongly in the volume emission regime on resonance, as discussed in the main text.

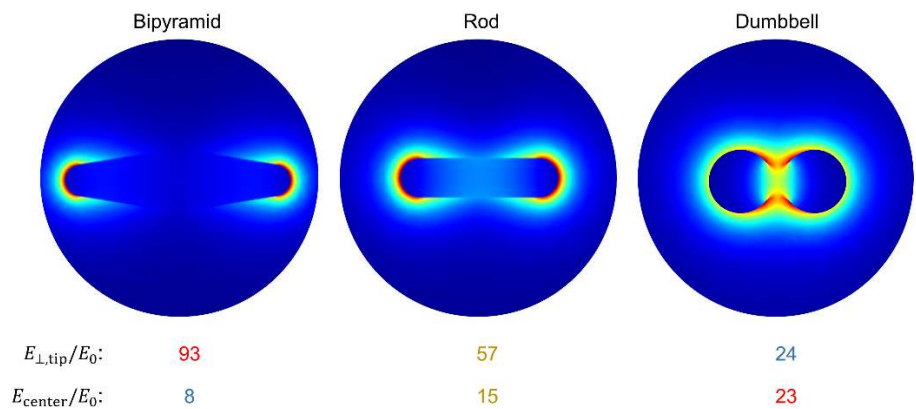

**Fig. S13.** Slices of plasmonic field enhancements for bipyramid, nanorod, and dumbbell geometries with the corresponding surface-normal tip enhancements and total volume center enhancements noted. All geometries are simulated in a vacuum environment with no surface ligands.

## Supporting References